\documentclass[12pt,a4paper]{article}
\usepackage[latin1]{inputenc}
\usepackage{amsmath}
\usepackage{amsfonts}
\usepackage{amssymb}
\usepackage{graphicx}
\usepackage{calc}
\usepackage{wrapfig}
\usepackage[a4paper,margin=1.0in]{geometry}
\usepackage{subfig}
\usepackage{hyperref}
\newcommand{\ee}{\ensuremath{\mathrm{e}^+\mathrm{e}^-}}
\newcommand{\ie}{{\it{\i.e.}}}
\newcommand{\eg}{{\it{e.g.}}}
\begin{document}
\title{\hfill{\normalsize {DESY 14-232}\\[-.4cm]\hfill{ November 2014}}\\[0.8cm]Photon collimator system for the ILC Positron Source}
\author{ S. Riemann, F. Staufenbiel\\\small\it{Deutsches Elektronen-Synchrotron DESY, 15738 Zeuthen, Platanenallee 6, Germany}\\
G. Moortgat-Pick, A. Ushakov\thanks{Work supported by the German Federal Ministry of Education and Research, Joint Research Project R\&D Accelerator ``Spin Optimization'', contract number 19XL7IC4} \\\small  \it{II. Institut f\"ur Theoretische Physik, Universit\"at Hamburg,
Luruper Chaussee 149,}\\ \small  \it{22761 Hamburg, Germany}
}
\date{}
\maketitle

\begin{abstract}
\noindent
High energy \ee   linear colliders are the next large scale project in  particle physics. They need intense sources to achieve the required luminosity. In particular, the  positron source must provide about 10$^{14}$ positrons per second. The positron source for the International Linear Collider (ILC) is based on a helical undulator passed by the electron beam to create an intense circularly polarized photon beam. With these photons a longitudinally polarized positron beam is generated; the degree of polarization can be enhanced by collimating the photon beam. However, the high photon beam intensity causes huge thermal load in the collimator material. In this paper the thermal load in the photon collimator is discussed and a flexible design solution is presented. 

\end{abstract}

\section{Introduction}

The positron source for the International Linear Collider (ILC)  is based on a helical undulator~\cite{ref:AccTDR}.
 Before collisions, the  accelerated electron beam passes the superconducting helical undulator and creates  an intense circularly polarized multi-MeV photon beam. The photons hit a positron target and create in an electromagnetic shower longitudinally polarized  positrons (and electrons). This method was suggested by Balakin and Mikhailichenko~\cite{ref:undulator-basics} and has been successfully tested with the E-166 experiment~\cite{ref:e166}. The baseline parameters of the ILC positron source afford a positron polarization of 30\%. The distribution of polarization within the photon beam depends on the radial position of the photons, so it is possible to increase the average polarization of positrons by  collimation from 30\% up to 50-60\%. 
However, 
the collimation of the photon beam  causes huge thermal load in the collimator material. In this paper, a photon collimator design is discussed which is based on studies of the  dynamic load in the collimator material. 
In section~\ref{sec:e+source} the ILC positron source is described, the photon collimator system is presented in section~\ref{sec:colli}. The thermal load as well as the cooling are discussed in section~\ref{sec:heatload+cool}; potential problems due to cyclic maximum load  and degradation are considered in section~\ref{sec:problems}. Finally, in section~\ref{sec:alternative} ideas for alternatives of the photon collimator design are presented which could overcame the drawback of the design presented here.

\section{ILC undulator based positron source for polarized positrons}\label{sec:e+source}

The ILC Technical Design Report (TDR)~\cite{ref:AccTDR} describes the machine parameters to get electron-positron collisions at centre-of-mass energies of 500\,GeV, 350\,GeV and 250\,GeV and also 1\,TeV. 
Trains of  1312 bunches (high luminosity option: 2625 bunches) with 2$\times$10$^{10}$ electrons/positrons  per bunch are repeated with a frequency of 5\,Hz.

\begin{figure}[h]
  \centering
  \includegraphics[width=150mm]{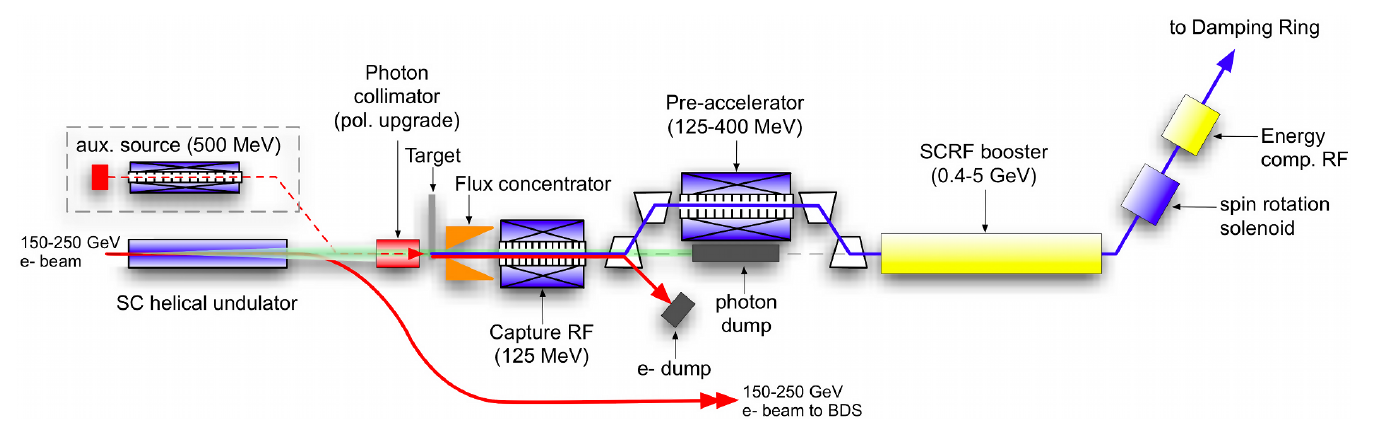}
  \label{ILC_target_wheel}
 \caption{Sketch of the positron production scheme for the ILC (see~\cite{ref:AccTDR}).}
\label{fig:source-sketch}
\end{figure}

The scheme of positron production is shown in figure~\ref{fig:source-sketch}. The superconducting helical undulator has a   period of $\lambda_0=11.5\,$mm and is  located  at a distance of 400\,m upstream the positron target.  
Depending on the electron beam energy and the desired polarization, 
the undulator K value varies  from $K=0.45$ up to $K=0.92$. 
The length of the undulator is determined by the requirement to generate  1.5 positrons per drive beam electron and  amounts up to 231\,m maximum.

The degree of photon polarization depends on the angular distribution of the photons. 
The intensity of the undulator radiation has the maximum around the beam axis. By cutting the outer part of the radial symmetric photon  beam with a collimator, the positron  polarization is increased by contemporaneous decreasing the   positron yield. 
The yield of 1.5\,e$^+$/e$^-$ can be  recovered by increasing the active length of the undulator and choosing $K = 0.92$.
Table~\ref{tab:e+pol}  illustrates 
the relation between undulator-K values, collimator aperture,  active length of the undulator and expected degree of  positron beam polarization using a flux concentrator as optical matching device with parameters described in the TDR~\cite{ref:AccTDR}. 
Depending on the electron beam energy and the K value, the positron polarization approaches  29\% for $r \geq 3\,$mm up to  50-60\% if the photon beam radii are collimated to $r=0.7-2.0\,$mm (see also \cite{ref:colli-pol} and table~\ref{tab:CollPar}). 
\begin{table}[h]
\begin{center} 
\renewcommand{\arraystretch}{1.14}
\begin{tabular}{|lc|c|c|c|c|}
\hline
 parameter                      & unit & \multicolumn{4}{|c|}{$E_\mathrm{cm}=500\,$GeV} \\\hline
\hline
$E_\mathrm{e^-}$                  & [GeV] &\multicolumn{4}{|c|}{ 250}  \\\cline{3-6}
K value                         & $-$  &   0.45     &  \multicolumn{3}{c|}{0.92}    \\\cline{3-6}
undulator length $L_\mathrm{und}$ &[m]   &   147      &  49   & 70  & 143.5\\
collimator iris radius          & [mm] &  $-$       & 1.4   & 1.0 & 0.7 \\
power absorption                & [kW] &  $-$       &  13   &  43 & 132 \\ 
$P_\mathrm{e^+}$                  & [\%] &   30       &  37   &  50 & 59 \\
\hline
\end{tabular}
\caption{Expected positron polarization, $P_\mathrm{e^+}$, for different undulator K values and photon collimator iris radii at $E_\mathrm{cm}=500\,$GeV, high luminosity. The active undulator length, $L_\mathrm{und}$, is adjusted to achieve the positron yield of 1.5\,e$^+$/e$^-$ for the nominal luminosity corresponding to 1312 bunches per train. The undulator period is $\lambda_\mathrm{und}=11.5\,$mm.}
\label{tab:e+pol}
\end{center}
\end{table}
Since the positron yield decreases with decreasing electron beam energy,    the so-called 10\,Hz scheme  has been proposed for centre-of-mass energies below 300\,GeV. It explores a 5\,Hz electron beam for physics alternating with another 5\,Hz electron beam of 150\,GeV to create the photons and subsequently the positrons.  
Current studies~\cite{ref:ushakov120GeV} show that also at low energies the electron beam could create enough photons to achieve the desired luminosity and 30\% positron polarization. An upgrade to 40\% is possible. However, a higher degree of polarization would require the 10\,Hz scheme. 
The study in this paper is aimed for  degrees of positron polarization of 50\% and higher. Therefore,  for $E_\mathrm{cm}=250\,$GeV the 10\,Hz scheme has been considered.

Some facts complicate the design of the photon collimator:
\begin{itemize}
\item  
The opening angle of the radiated photon beam is determined by the energy of the electron beam; it is proportional to $1/\gamma$.
Although the helical undulator is located at a distance of 400\,m upstream the positron target, the photon beam spot is small. The energy deposition density along the path of the intense photon beam is large  in the collimator and the positron target. The conversion target is designed as spinning wheel, so the thermal load is substantially reduced. 
The photon collimator is fixed, static and it  has to stand  a huge power absorption. 
Depending on the electron beam energy and the desired positron polarization, the  average photon beam power is in the range of $P_\gamma=83-340\,$kW (see also table~\ref{tab:CollPar}). 
\item 
Since the positron source is located at the end of the main linac, its parameters  are  strongly coupled to the centre-of-mass energy of the collider. This also applies to the photon collimator. It is impossible to cover with one design the requirements for all centre-of-mass energies.
\end{itemize}

\section{ILC photon collimator system}\label{sec:colli}

Three important energies are considered for running the ILC: 
\begin{itemize}
\item $E_\mathrm{cm} = 250\,$GeV to produce Higgs bosons, mainly by the Higgs-Strahlungs process,
\item $E_\mathrm{cm} = 350\,$GeV to study the threshold of top-quark pair production,
\item $E_\mathrm{cm} = 500\,$GeV, the nominal energy to study interesting processes of the Standard Model and beyond.
\end{itemize}
For the baseline option, the undulator parameters are adjusted to get a positron beam which is 22\% up to about 30\% polarized. 
With a photon collimator the positron polarization can be increased: The aperture of the collimator determines the average polarization of the photon beam at the target and hence, of the positrons produced and captured. 

To achieve flexibility in polarization and yield manipulation for the positron beam which is coupled to the energy of the electron beam,  a 
 system of three collimators  with attenuating apertures  is proposed.  
To increase the positron polarization up to 50-60\%, 
at  $E_\mathrm{cm}=250\,$GeV only the  first collimator is needed, at  $E_\mathrm{cm}=350\,$GeV the first and second collimators  and at   $E_\mathrm{cm}=500\,$GeV all three collimators are used. 

In the past, several possibilities for a collimator design were discussed~\cite{ref:colli_alexander,ref:colli_lei-zang} but a  flexible design suitable for  different centre-of-mass energies was not considered in detail. 
All collimator designs have to deal with the intense, focused photon beam. Simple, adjustable spoilers in front of  the absorber material are less effective for a photon than for an electron beam.

The collimator design described here consists of stationary parts, \ie\ the collimator material is not moved to distribute the heat load over a 
larger volume.
A collimator system with moving (rotating) components is also possible and currently under consideration. 

In order to choose material and dimensions of the collimator, the electromagnetic shower distribution and the corresponding energy deposition in the material has to be considered.

\subsection{Electromagnetic shower in the collimator}\label{sec:shower}
The energy of the undulator photons (first harmonic) is  few MeV up to few tens MeV depending on the drive electron beam energy,  the opening angle of the photon beam is proportional to $ 1/\gamma_\mathrm{e-}$.
The photon collimator should absorb the outer part of the photon beam. 
At high energies, the characteristic interaction length 
of photons is given by the radiation length $X_0$. %
In principle,    a high Z-material with large density and small radiation length would convert the photons and stop the remaining particles best so that the collimator could be quite compact. 
However, a shorter radiation length corresponds to a higher  pair-production cross section, and so 
the density of the produced shower particles, \ie\ e$^+$e$^-$ pairs and Bremsstrahlungs photons, is enhanced  and the energy deposition density increases. 
That means that the temperature rise could be too large in the critical region at and  near the inner surface of the collimator.

To choose a reliable design of the collimator, the passage of the photon beam through the collimator material has been simulated. The dimension and shape of the collimator as well as the material were adjusted by keeping the temperature rise along z-direction at an acceptable, relatively constant level, {\it{e.g.}}\ to avoid sharp temperature jumps. This reduces the stress in the collimator material and prevents overloading.

\subsection{Basic  collimator layout}\label{sec:layout}
The collimator design  suggested consists of three parts; 
the minimal aperture radii are  2\,mm, 1.4\,mm and 1.0\,mm for the first, second and third collimator. 
With the first collimator the polarization is increased up to 50\% for electron beam energies of 150\,GeV ($E_\mathrm{cm}=250\,$GeV). The first and the  second or all three collimators are necessary to achieve positron  polarization above 50\% for centre-of-mass energies of $E_{\mathrm{e^-}} = 350\,$GeV or $E_{\mathrm{e^-}} = 500\,$GeV, respectively. 

\begin{figure}[h]
  \centering
  \includegraphics[width=125mm]{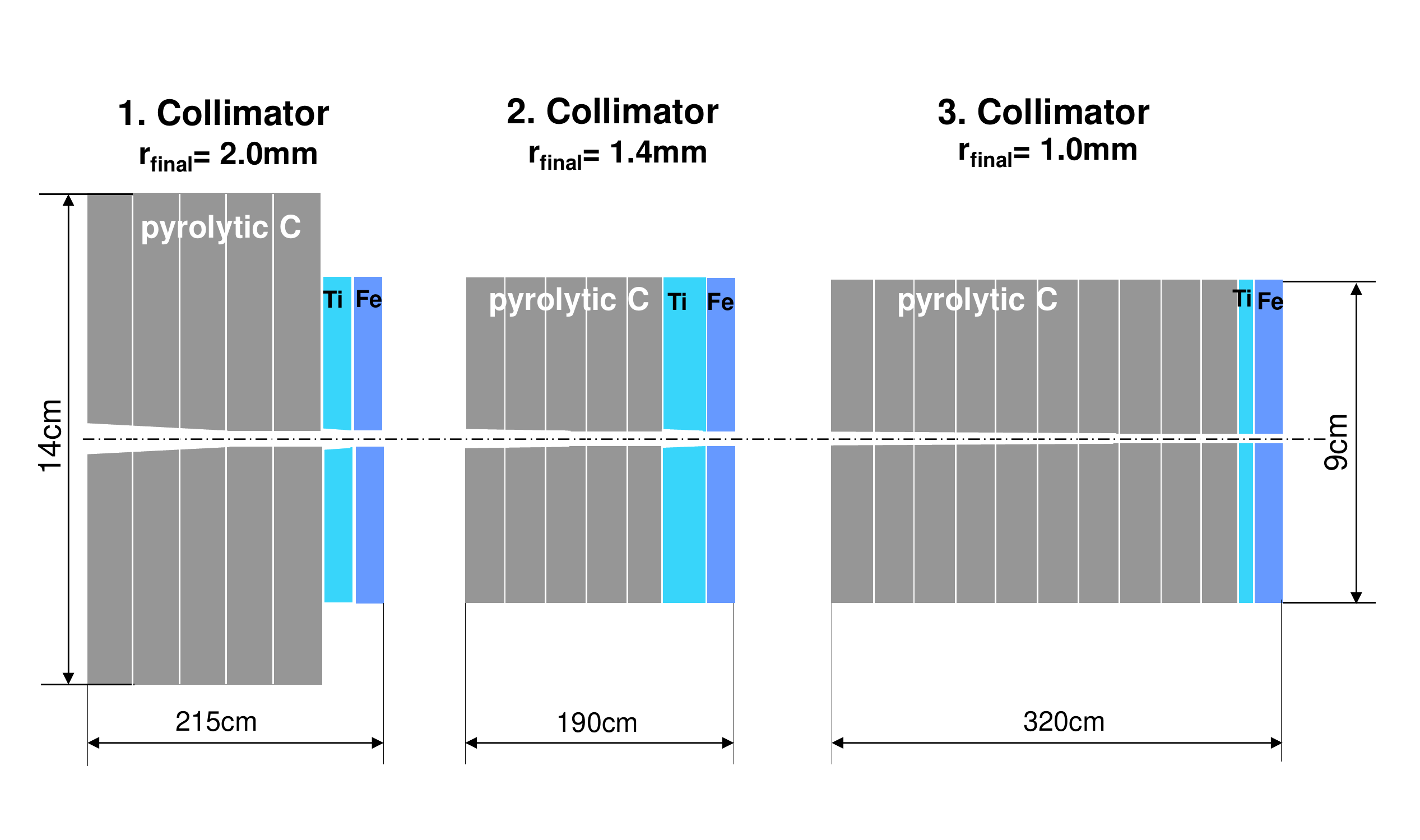}
  \caption{Sketch of the multistage collimator system. For details see also table~\ref{tab:gCollDim}. 
}
  \label{fig:Collimators_basic-layout1}
\end{figure}

The  energy deposition in the  collimator has been 
calculated using the FLUKA Monte Carlo code for particle tracking and particle interactions with matter \cite{ref:FLUKA}. By means of this simulation tool the optimization of the photon collimator design is done by quantifying the heat load in the collimator and selecting the material corresponding to the tolerable thermo-mechanical stress. 
The temperature rise, $\mathrm{d} T$, is described by
\begin{equation}
 \mathrm{d} Q=m\,c\, \mathrm{d}T 
\label{eq:DeltaT}
\end{equation}
 where $Q$ is the energy deposition in the material,  
$m$ the mass and $c$ is the specific heat capacity. 
The volume elements with highest energy deposition, the so-called  peak energy deposition density (PEDD), experience the maximal temperature rise.
The collimator design has to avoid PEDD values which could  damage the material and cause failure of the collimator. 
The results of calculations and simulations resulted in a system which is sketched in figure~\ref{fig:Collimators_basic-layout1} and presented in the following sections. 

\subsubsection{Collimator material}\label{sec:material}

The main fraction of energy is absorbed in the first part of each collimator stage. A low-Z material, pyrolytic graphite, has been chosen. Its evaporation point scores up to 3650$^\circ$C without a liquid phase \cite{ref:J.Pappis}. In addition,pyrolytic graphite is very resistant against particle evaporation by energy impact. The material is strong anisotropic in the (xy) plane (basal direction) and in (z) direction; the thermal conductivity is a factor of 200 higher in the basal direction and has a very low thermal expansion coefficient \cite{ref:D.Yao}. However, due to the high radiation length of $X_0 \approx 19\,$cm, a very long collimator is needed to absorb the whole unwanted part of the photon beam. In order to distribute the energy deposition 
in the collimator material and to keep the collimator as short as possible, proper medium-Z (or high-Z) material with smaller radiation length has to follow the graphite segments. Titanium alloy (Ti8Mn)  and iron (St-70) have been chosen as  collimator material behind the pyrolytic graphite parts. 
\begin{table}[h]
 \renewcommand{\arraystretch}{1.14}
\begin{tabular}{|lc|cc|c|c|}
\hline
 parameter                    &unit                    & \multicolumn{2}{|c|}{pyr. C} & Ti8Mn &  Iron (St-70)\\
                              &                        &(x,y)  & (z)                  &         &            \\
\hline
\hline
density  $\rho$                &g/cm$^3$                &\multicolumn{2}{|c|}{ 2.2}  &   4.7 &  7.9 \\
specific heat capacity $c$    &J/$(\mathrm{g \cdot K})$&\multicolumn{2}{|c|}{0.837} & 0.495 &  0.434 \\
thermal conductivity $\lambda$&W/$(\mathrm{m \cdot K})$&   346       &         1.73 & 11--16&  61  \\
coeff. of thermal expansion  $\alpha$&$10^{-6}/\mathrm{K}$&       0.5 &         6.5  &    10 &  12     \\
critical energy $E_\mathrm{C}$  & MeV                    &\multicolumn{2}{|c|}{81.7} & 26.0&  21.8          \\
radiation length $X_0$         &cm                     &\multicolumn{2}{|c|}{19.32} & 3.56 & 1.76         \\
melting point                 & K                      &\multicolumn{2}{|c|}{3900}  & 1565  & 1870 \\
                              &                        &\multicolumn{2}{|c|}{(sublimation)} &  &      \\
modulus of elasticity $Y$     & GPa                    &\multicolumn{2}{|c|}{20}    &  115  &  200   \\
Poisson's ratio  $\nu$        &                        & $-0.1$      &   0.3        &  0.33 &  0.3    \\
ultimate tensile strength     & MPa                    &\multicolumn{2}{|c|}{90}    &  900  &  700 \\
tensile strength (yield)      & MPa                    &\multicolumn{2}{|c|}{90}    &  810  &  340 \\
fatigue strength              & MPa                    &\multicolumn{2}{|c|}{$ $}   &  $210-585$  &  280 \\
\hline
\end{tabular}
\caption{Parameters of the material used in the collimator (see also~\cite{ref:MatWeb,ref:Dubbel,ref:pdg}). 
}
\label{tab:matpar}
\end{table}

\begin{figure}[hp]
  \centering
  \includegraphics[width=100mm]{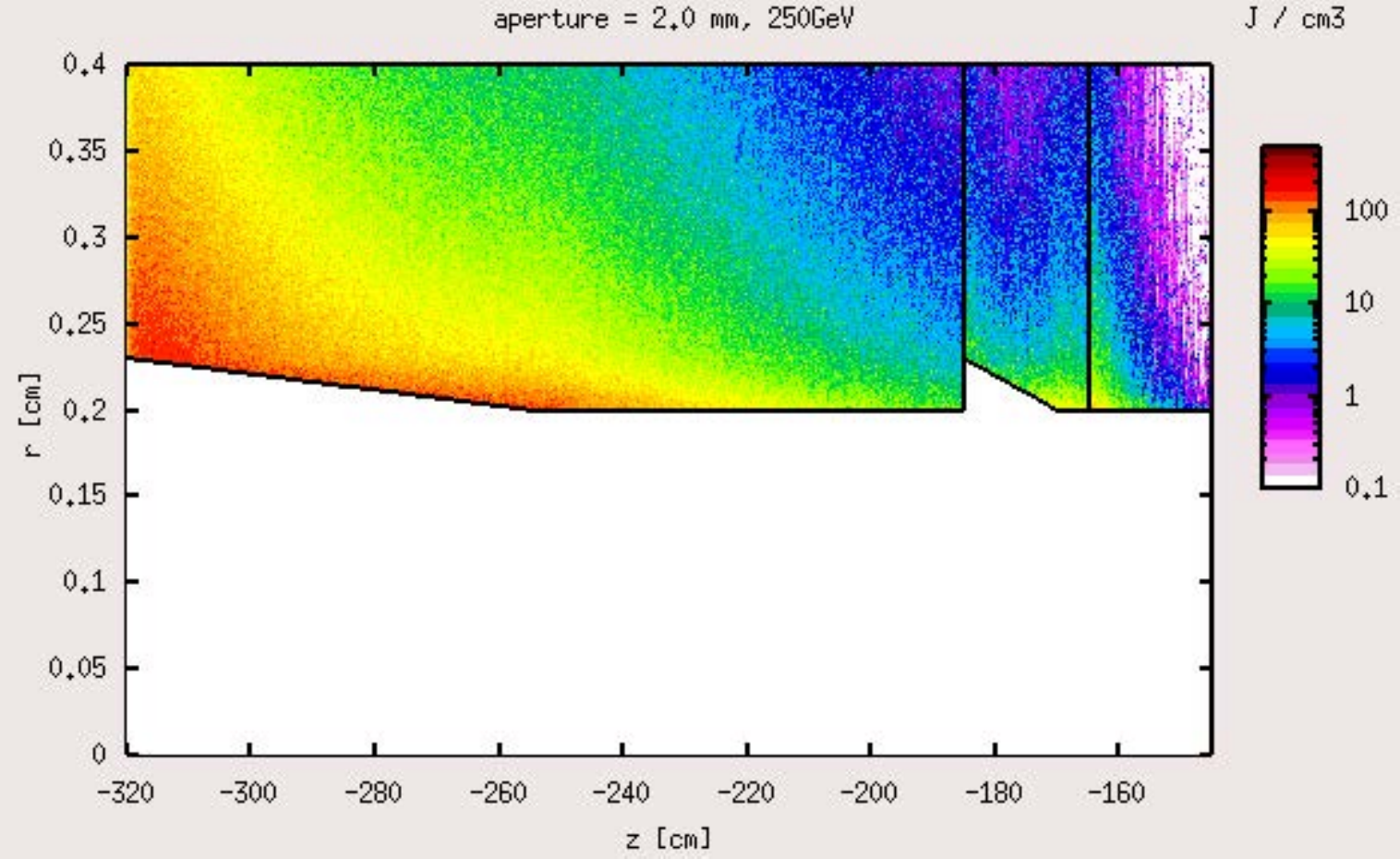}\\
  \includegraphics[width=100mm]{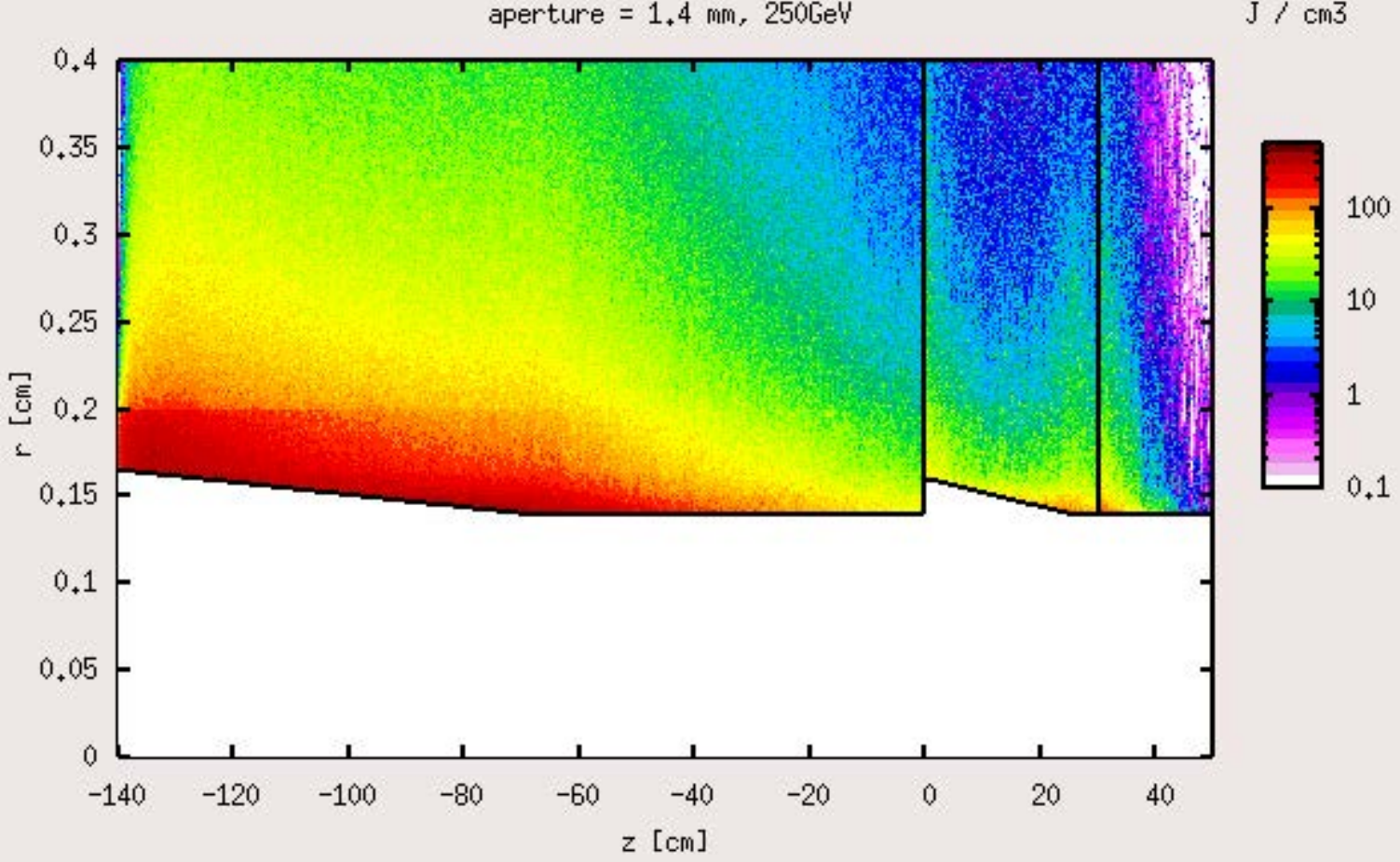}\\
  \includegraphics[width=100mm]{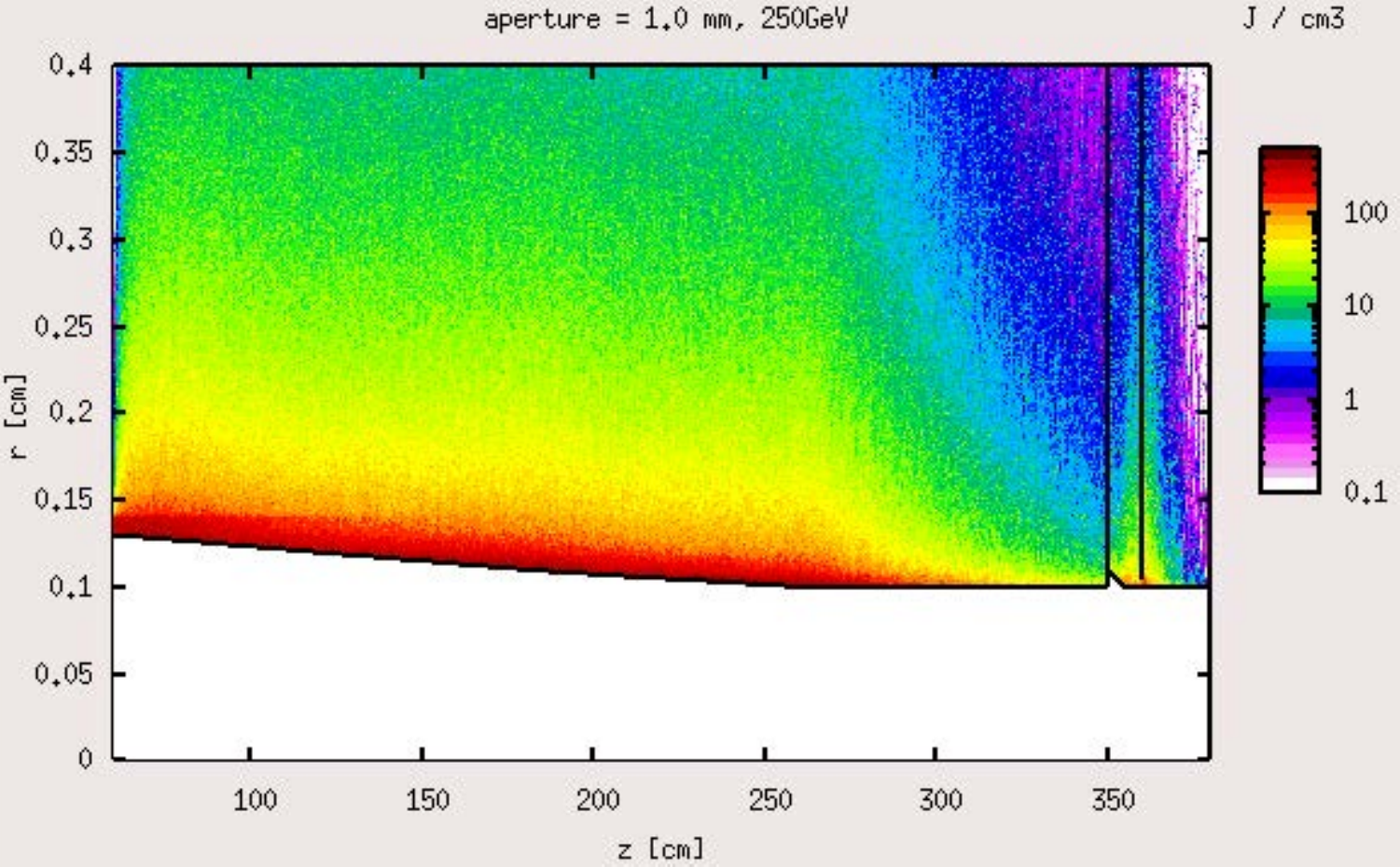}
  \caption{Distribution of deposited energy in the photon collimator. Shown are the three stages with decreasing aperture radii $r_\mathrm{min}= 2$\,mm, 1.4\,mm and 1.0\,mm for an electron beam energy $E\mathrm{e^-}= 250\,$GeV. 
}
  \label{fig:ILC_collimatorR12}
\end{figure}

Figure~\ref{fig:ILC_collimatorR12} shows the simulated energy deposition distributions in all three collimator stages assuming a 250\,GeV electron beam to achieve 50\% positron polarization. It is clearly visible that the collimator design allows a quite uniform temperature distribution along the collimator aperture. 

It should be remarked that for the simulations the collimator segments are assumed as simple, one-piece blocks.
The fabrication of such segments with small aperture including cooling channels has not been regarded; most likely the longer collimator components will consist of partitioned segments. The proper alignment of the collimators with small aperture segments requires special care to obtain the desired reproducibility of polarization.

\subsubsection{Dimensions of the first, second and third collimator}\label{sec:dimensions}

The lengths of the collimator parts are optimized to lower the energy deposited in the following higher Z material to an acceptable level. 
Since the average energy of the photons is  below or near the critical energy  of the collimator material, the highest temperatures occur at and near the inner surface. 
 In order to distribute the load over a larger volume in the collimator material, slightly tapered sections are inserted  at the beginning of the graphite and titanium parts. In case of the first and third collimator (2\,mm and 1\,mm final iris radius), the graphite part is tapered in two steps (see table~\ref{tab:gCollDim}).
In the iron part the aperture is not tapered.

At higher centre-of-mass energies smaller collimator apertures are required to achieve high positron polarization. Since the photon cut-off energy increases with the squared energy of the electrons passing the undulator, 
also a longer collimator is required to stop the photons. 
In particular, the pyrolytic part must be longer  to protect the following titanium and iron sections. 

\begin{figure}[h]
  \centering
  \includegraphics*[width=0.95\textwidth]{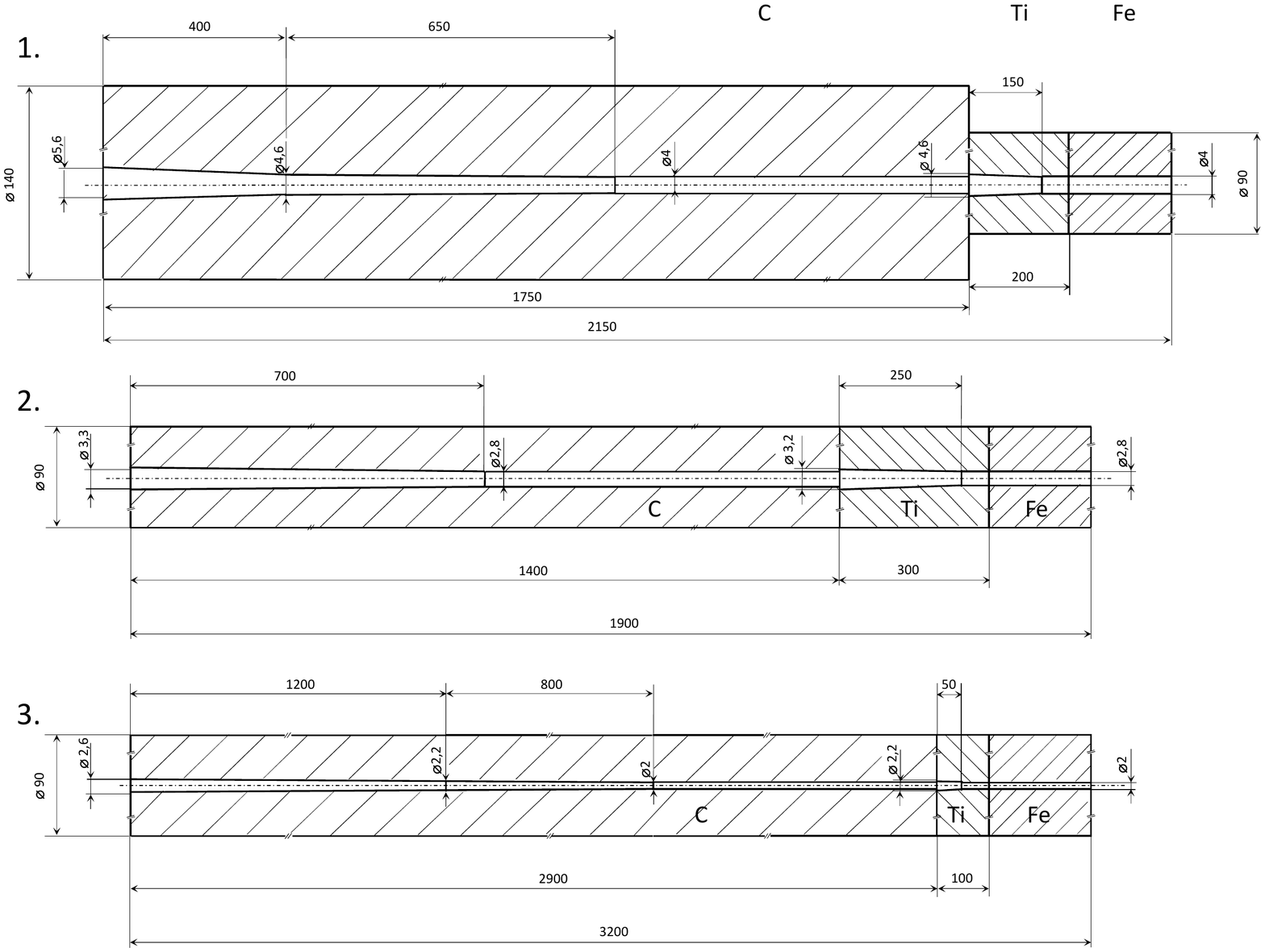}
  \caption{Technical drawing  of the collimator main parts; the cooling is not included. 
}
  \label{fig:Collimators_sketch}
\end{figure}

The dimension given in table~\ref{tab:gCollDim} and  shown in figure~\ref{fig:Collimators_sketch} represent the suggested collimator design. It allows to achieve almost 60\% positron polarization at $E_\mathrm{cm}=350\,$GeV and 50\% at $E_\mathrm{cm}=500\,$GeV.  
In addition, the study included the possibility to achieve  60\% positron polarization at $E_\mathrm{cm}=500\,$GeV.  In this case all apertures of each collimator was reduced by 0.3\,mm keeping length and outer diameter as given in   table~\ref{tab:gCollDim};
However, in this case the power dumped in the photon collimator would be almost three times as much as in the case of 50\% positron polarization (see also table~\ref{tab:CollPar}). 
\begin{table}[h] 
 \renewcommand{\arraystretch}{1.14}
\begin{flushleft}
\begin{tabular}{|l|ccc|ccc|ccc|  }  \hline 
                        &  \multicolumn{3}{c|}{1. collimator }
                        & \multicolumn{3}{c|}{2. collimator}
                        & \multicolumn{3}{c|}{3. collimator} \\\hline
   final                & \multicolumn{3}{c|}{  }
                        & \multicolumn{3}{c|}{  }
                        & \multicolumn{3}{c|}{  } \\  
   iris radius          & \multicolumn{3}{c|}{2\,mm}
                        & \multicolumn{3}{c|}{ 1.4\,mm}
                        & \multicolumn{3}{c|}{1\,mm} \\ \hline   
    &{\footnotesize length} &{\footnotesize out. rad. }&{\footnotesize weight}
    &{\footnotesize length} &{\footnotesize out. rad.}&{\footnotesize weight}
    &{\footnotesize length} &{\footnotesize out. rad.}&{\footnotesize weight}\\
    &{\footnotesize  [mm]}  &{\footnotesize  [mm]}&{\footnotesize  [kg]}
    &{\footnotesize  [mm]}  &{\footnotesize  [mm]}&{\footnotesize  [kg]} 
    &{\footnotesize  [mm]}  &{\footnotesize  [mm]}&{\footnotesize  [kg]} 
\\ \cline{2-10}
    pyr. C      & 1,750  & 70 &  72.7 & 1,400 & 45 & 31.2 &  2,900 & 45 & 64.6 \\  
    Ti8Mn       &  200   & 45 &  5.7  &   300 & 45 &  8.6 &    100 & 45 &  2.9 \\
    Iron        &  200   & 45 &  10   &   200 & 45 &   10 &    200 & 45 &   10  \\%
 \hline
   {\footnotesize active length} & 2,150 &  &      & 1,900 & &     & 3,200 & & \\\hline
 \multicolumn{10}{|l|}{taper}\\
 \multicolumn{10}{|l|}{parameters  }\\ \hline
                &{\footnotesize length} &\multicolumn{2}{c|} {\footnotesize $r_1 ~/~ r_2$}
                &{\footnotesize length} &\multicolumn{2}{c|} {\footnotesize $r_1 ~/~ r_2$}
                &{\footnotesize length} &\multicolumn{2}{c|} {\footnotesize $r_1 ~/~ r_2$} \\ 
                &{\footnotesize [mm]} &\multicolumn{2}{c|} {\footnotesize [mm / mm]}
                &{\footnotesize [mm]} &\multicolumn{2}{c|} {\footnotesize [mm / mm]}
                &{\footnotesize [mm]} &\multicolumn{2}{c|} {\footnotesize [mm / mm]} \\ \hline
  {\footnotesize  pyr. C }   &  400 &\multicolumn{2}{c|} {2.9 / 2.3}   &  700 &\multicolumn{2}{c|} {1.65 / 1.4}  & 120 &\multicolumn{2}{c|}{ 1.3 / 1.1}   \\
                             &  650 &\multicolumn{2}{c|} {2.3 / 2.0}   &     & \multicolumn{2}{c|} { }           & 800 &\multicolumn{2}{c|}{ 1.1 / 1.0}  \\
 {\footnotesize  Ti8Mn }     &  150 &\multicolumn{2}{c|} {2.3 / 2.0}   &  250 &\multicolumn{2}{c|} {1.60 / 1.4}  &  50 &\multicolumn{2}{c|}{1.1 / 1.0}   \\
    \hline
\end{tabular}
\caption[Dimensions of the e+ source photon collimator parts.]{\label{tab:gCollDim}Dimensions of the photon collimator parts. The tapered sections are described by the aperture radius $r_1$ at the beginning and $r_2$ at the end.} 
\end{flushleft}
\end{table}

The collimator absorbs more than 99.9\% of the unwanted part of the photon beam and the secondary particles; less than 0.1\% reaches the positron production target.

\section{Thermal load and cooling of the collimator}\label{sec:heatload+cool}
The average power and the peak energy deposited in the collimator, $P_\mathrm{ave}$ and $E_\mathrm{max}$, as well as the absorbed power have been simulated with FLUKA and ANSYS. They are summarized in table~\ref{tab:CollPar} for the different centre-of-mass energies and degrees of positron polarization.  These values are used to calculate the requirements for the cooling system and to evaluate the thermo-mechanical load in the collimator. 
The numbers in table~\ref{tab:CollPar} figure out that 
about half of the photon beam power is dumped in the collimator to get 50\% positron polarization. In order to reach  60\% positron polarization at $E_\mathrm{cm}=500\,$GeV, the power absorption rises to 75\% with a final collimator iris radius of  0.7\,mm.  
\begin{table}[p]
 \renewcommand{\arraystretch}{1.14}
 \centering
{\footnotesize  
\begin{tabular} {|llc||c|c||c||c|c|c|c|c|} \hline
\multicolumn{3}{|l||}{photon collimator parameters} & \multicolumn{8}{|c|}{cms energy [GeV]} \\ \cline{4-11}
\multicolumn{3}{|l||}{ }                            &  \multicolumn{2}{|c||}{250} & 350 & 500 &\multicolumn{4}{|c|}{500 (high lumi)}  \\ \hline \hline
\multicolumn{2}{|l}{electron beam energy $E_\mathrm{e-}$}& [GeV] &150&125 &  178 &  253 &\multicolumn{4}{c|}{253} \\ \hline
\multicolumn{2}{|l}{repetition rate}&[Hz]                   & 5 & 5 &   \multicolumn{6}{|c|}{5} \\ \hline
\multicolumn{3}{|l||}{number of e$^+$ bunches}              & \multicolumn{4}{|c|}{1312} & \multicolumn {4}{|c|}{2625}   \\ \hline
\multicolumn{2}{|l}{active undulator length}& [m] & 231 & 192.5 &  196 &  70 &\multicolumn{2}{c|}{70}   &\multicolumn{2}{c|}{143.5} \\ \hline
\multicolumn{2}{|l}{photons / train}& [x 10$^{15}$] & 11.8 & 9.8 & 10.0 & 3.6 &\multicolumn{2}{c|}{7.2}  &\multicolumn{2}{c|}{14.6} \\ \hline
\multicolumn{2}{|l}{average photon power}& [kW]   & 98.5 & 68.4 &113.6 &82.9 &\multicolumn{2}{c|}{166.2}&\multicolumn{2}{c|}{339.5} \\ \hline
\multicolumn{2}{|l}{1st harmonic cut-off}& [MeV]  &  10.1    &  7.0    &   14.2   &  \multicolumn{5}{c|}{ 28.6  }\\ \hline
\multicolumn{2}{|l}{final iris radius}& [mm]& 2.0 & 2.0 & 1.4 & 1.0 &\multicolumn{2}{c|}{1.0}  &\multicolumn{2}{c|}{0.7}  \\ \hline
\multicolumn{2}{|l}{e$^+$ polarization}& [\%]     &55.3 & $-$ & 58.5 &50.3 &\multicolumn{2}{c|}{50.3} &\multicolumn{2}{c|}{58.7} \\ \hline
\multicolumn{2}{|l}{transversal mismatch}& $\Delta x\,[\mu m]$ & \multicolumn {4}{|c|}{$-$}&   0  & 100  & 0     & 100  \\ \hline
\multicolumn{2}{|l}{absorbed power in collimator}& [kW]        & 48.5& 42.4 & 68.7 &  43.5   & 87 & 87.3  & 254.8 & 255  \\ \hline \hline
\multicolumn{3}{|l||}{1. collimator }
& \multicolumn{6}{|c|}{ final iris radius $2$\,mm }& \multicolumn{2}{c|}{1.7\,mm} \\  \hline
Pyr. C:& $E_\mathrm{max}$ &[J/g]        &  53  &  45 &  53  &  13 &   26 &  33  &   99 &  129 \\
   &$\Delta T_\mathrm{max}$&[K]         &  63  &  54 &  63  &  16 &   31 &  39  &  118  & 154 \\
   &$P_\mathrm{ave}$       & [kW]       & 45.2 &  40 & 36.3 & 7.9 & 15.8 & 16.0 & 52.8 & 53.2 \\ \hline
Ti:& $E_\mathrm{max}$ &[J/g]            &   10 &  5  &  10  &   2 &    5 &   10 &   23 &  39  \\
   &$\Delta T_\mathrm{max}$& [K]        &   20 &  10 &  20  &   4 &   10 &   20 &   46 &  79  \\
   &$P_\mathrm{ave}$       & [kW]       &  0.6 & 0.3 &  0.8 & 0.2 &  0.4 &  0.4 &  2.0 &  2.0 \\ \hline
Fe:& $E_\mathrm{max}$& [J/g]            &    7 &   4 &   7  &   2 &    4 &   7  &   18 &  28  \\
   &$\Delta T_\mathrm{max}$ &[K]        &   16 &   9 &  16  &   5 &    9 &   16 &   42 &  65  \\
   &$P_\mathrm{ave}$ &[kW]              &  0.3 & 0.1 &  0.3 & 0.1 &  0.2 &  0.2 &  0.9 &  0.8 \\ \hline
Cu:& $P_\mathrm{ave}$& [kW]             &  2.4 & 2.0 &  1.9 & 0.4 &  0.8 &  0.8 &  2.8 &  2.9 \\ \hline \hline
\multicolumn{5}{|l||}{2. collimator  }& \multicolumn{4}{|c|}{ final iris radius $1.4$\,mm } & \multicolumn{2}{c|}{1.1\,mm} \\  \hline 
Pyr. C: & $E_\mathrm{max}$ & [J/g]      & \multicolumn{2}{c||}{$-$} &  104 &   40 &   81 & 100  &  318 & 408 \\
   &$\Delta T_\mathrm{max}$& [K]        & \multicolumn{2}{c||}{$-$} &  124 &   49 &   97 & 119  &  380 & 488 \\
   &$P_\mathrm{ave}$ &[kW]              & \multicolumn{2}{c||}{$-$} & 25.9 & 12.9 & 25.8 & 26.0 & 82.6 & 83.1\\ \hline
Ti:&$E_\mathrm{max}$ &[J/g]             & \multicolumn{2}{c||}{$-$} &   15 &    9 &  18  &   26 &   72 &  98  \\
   &$\Delta T_\mathrm{max}$ &[K]        & \multicolumn{2}{c||}{$-$} &   30 &   18 &  36  &   53 &  145 & 198 \\
   &$P_\mathrm{ave}$ &[kW]              & \multicolumn{2}{c||}{$-$} &  0.9 &  0.6 &  1.2 &  1.2 &  4.1 & 4.1 \\ \hline
Fe:&$E_\mathrm{max}$ &[J/g]             & \multicolumn{2}{c||}{$-$} &   11 &    6 &   13 &   18 &   48 & 71 \\ 
   &$\Delta T_\mathrm{max}$ &[K]        & \multicolumn{2}{c||}{$-$} &   25 &   14 &  30 &    42 &  111 & 164 \\
   &$P_\mathrm{ave}$ &[kW]              & \multicolumn{2}{c||}{$-$} &  0.2 &  0.2 &  0.3 &  0.3 &  1.1 & 1.1 \\ \hline
Cu:& $P_\mathrm{ave}$& [kW]             & \multicolumn{2}{c||}{$-$} &  2.3 &  1.2 &  2.4 &  2.5 &  8.2 & 8.2 \\ \hline \hline
\multicolumn{6}{|l||}{3. collimator  }& \multicolumn{3}{|c|}{final iris rad. 1\,mm}&\multicolumn{2}{c|}{0.7\,mm} \\  \hline
Pyr. C:& $E_\mathrm{max}$& [J/g]        & \multicolumn{3}{c||}{$-$} &   47 &   95 &  120 &  325 & 377  \\
   &$\Delta T_\mathrm{max}$&[K]         & \multicolumn{3}{c||}{$-$} &   56 &  113 &  143 & 388  & 450  \\
   &$P_\mathrm{ave}$ &[kW]              & \multicolumn{3}{c||}{$-$} & 17.9 & 35.8 & 35.8 & 90.0 & 89.5 \\ \hline
Ti:& $E_\mathrm{max}$& [J/g]            & \multicolumn{3}{c||}{$-$} &   10 &   19 &   32 &   65 & 86  \\
   &$\Delta T_\mathrm{max}$& [K]        & \multicolumn{3}{c||}{$-$} &   20 &   38 &   64 &  131 & 174 \\
   &$P_\mathrm{ave}$ &[kW]              & \multicolumn{3}{c||}{$-$} &  0.1 &  0.3 & 0.3  &  0.6 & 0.6 \\ \hline
Fe:& $E_\mathrm{max}$&[J/g]             & \multicolumn{3}{c||}{$-$} &    7 &   15 &   24 &  50  &  63 \\
   &$\Delta T_\mathrm{max}$&[K]         & \multicolumn{3}{c||}{$-$} &   16 &   35 &   55 &  115 & 145 \\
   &$P_\mathrm{ave}$  &[kW]             & \multicolumn{3}{c||}{$-$} &  0.1 &  0.3 &  0.3 &  0.5 & 0.6 \\ \hline
Cu:& $P_\mathrm{ave}$ &[kW]             & \multicolumn{3}{c||}{$-$} &  1.8 &  3.7 &  3.7 &  9.2 & 9.1 \\ \hline
\end{tabular} 
}
  \caption{\label{tab:CollPar}Maximum energy deposition, $E_\mathrm{max}$, and  maximum temperature increase, $\Delta T_\mathrm{max}$, by one bunch train, and average power deposition, $P_\mathrm{ave}$, in the collimator parts  for different centre-of-mass energy options. 
The undulator parameters are $K=0.92$ and $\lambda_0 = 11.5\,$mm. The positron yield is 1.5 e$^+$/e$^-$.
}
\end{table}

\subsection{Collimator cooling system}\label{sec:coolsystem}
In the equilibrium, the radial heat dissipation through a hollow cylinder with central heating is given by \cite{ref:W.Wagner}  
\begin{equation}
\displaystyle{\frac{\mathrm{d}Q}{\mathrm{d}t}}=\displaystyle{\frac{ 2\pi\lambda z\Delta T}{\ln(\frac{r}{r_0})}}
\label{eq:dQdtcyl}
\end{equation}
where $r_0$ and $r$ are the inner and outer radius of the cylinder, $z$ its length, and $\lambda$ the thermal conductivity. Equation~(\ref{eq:dQdtcyl}) is used to adjust the outer radius of the collimator and the cooling power required to achieve the average temperature difference $\Delta T$ between inner and outer surface of the cylinder. 
Assuming a homogeneous, radial directed thermal dissipation from the inner hot to the outer cooled surface, the required average cooling power corresponds to $\dot{Q}$.

Due to the slightly conical apertures the maximum heat load is distributed  and kept within reasonable limits.  
 The time-dependent temperature rise and fall with each bunch train smears out in the bulk of the collimator since the heat transfer from the inner to the outer surface takes few seconds. 
The average heat flux through the outer surface of the collimator parts -- mainly the graphite -- is below 10\,W/cm$^2$.

\begin{figure}[hp]
  \centering
  \includegraphics*[width=100mm]{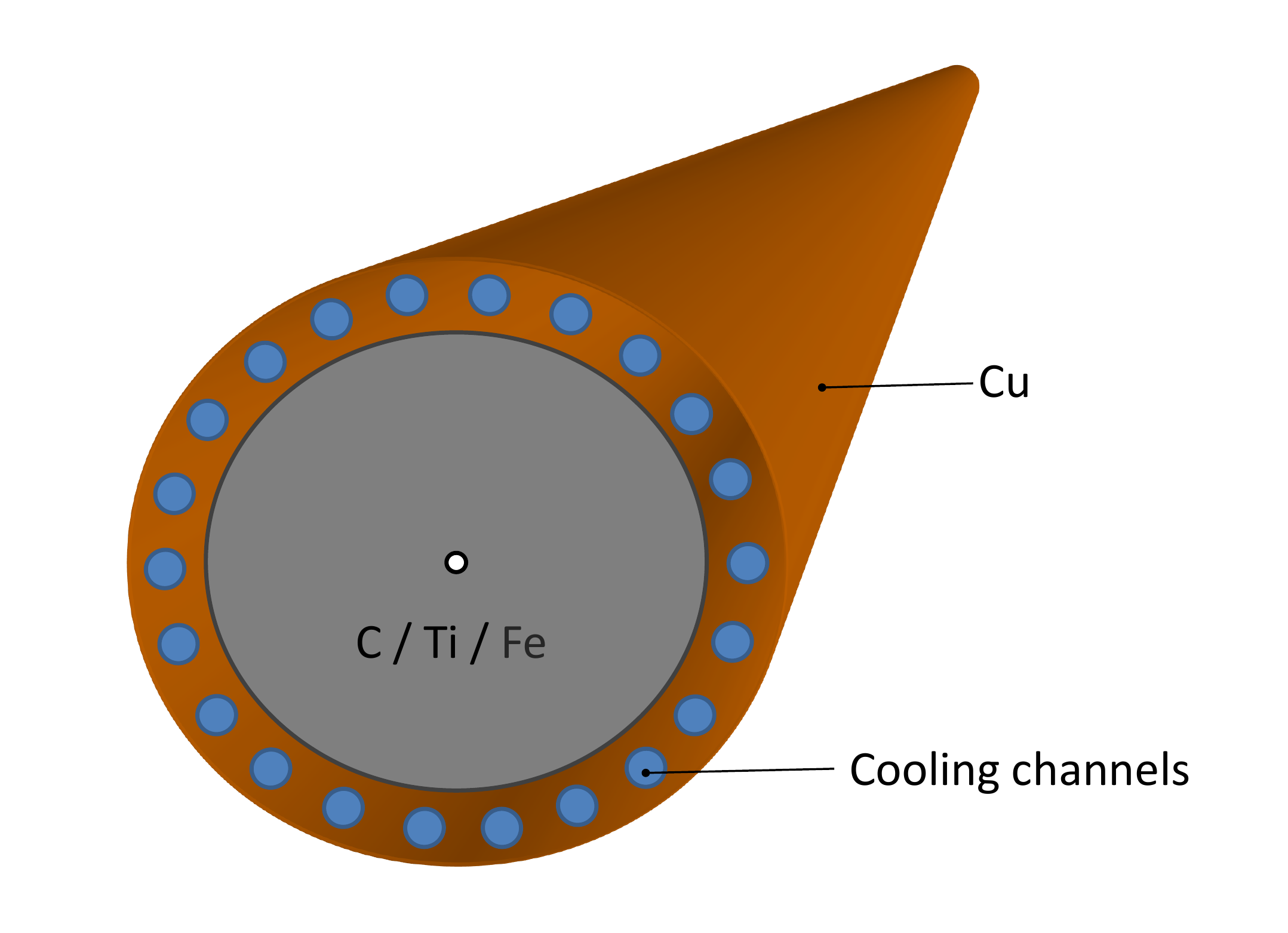}
  \caption{Collimator sketch with cooling channels. The amount of channels depends on the outer radius of the collimator which corresponds to the total cooling power.}
  \label{fig:cool_collimator}
\end{figure}

 For a technical solution of cooling the collimators are jacketed with copper, a material of high thermal conductivity.  Straight cooling channels are embedded in  2\,cm copper as shown in figure~\ref{fig:cool_collimator}. Due to the dimension chosen for  the collimator material graphite, titanium and iron,  the photon beam is stopped in these material and only a small part of the shower tail reaches the copper layer. The total power deposited in the copper is listed in table~\ref{tab:CollPar}.

The cooling water has to absorb the power given in equation~(\ref{eq:dQdtcyl});
\begin{equation}
\frac{\mathrm{d} Q_\mathrm{W}}{\mathrm{d}t} = \frac{\mathrm{d} Q_\mathrm{coll}}{\mathrm{d}t} = \frac{\mathrm{d} m_\mathrm{W}}{\mathrm{d}t}c_\mathrm{W}\Delta T_\mathrm{W}
 = \frac{\mathrm{d} V_\mathrm{W}}{\mathrm{d}t}\rho_\mathrm{W}c_\mathrm{W}\Delta T_\mathrm{W}\,,
\label{eq:cool1}
\end{equation}
where 
$\Delta T_\mathrm{W}$ is the average difference  between incoming and out-coming water temperature in the cooling tubes, and  $\mathrm{d} V_\mathrm{W}/{\mathrm{d}t}$ corresponds to the water flow needed to carry away   ${\mathrm{d} Q_\mathrm{coll}}/{\mathrm{d}t}$.
With $N$ cooling channels of radius $r_\mathrm{W}$ one gets
\begin{equation}
\frac{\mathrm{d} Q_\mathrm{W}}{\mathrm{d}t} = N \frac{\mathrm{d} z_\mathrm{W}}{\mathrm{d}t}  \pi r^2_\mathrm{W} \rho_\mathrm{W}c_\mathrm{W}\Delta T_\mathrm{W}\,,
\label{eq:cool2}
\end{equation} 
which  allows to determine  the required  velocity of the cooling water, $v_\mathrm{W}$;
 \begin{equation}
 \frac{\mathrm{d}z_\mathrm{W}}{\mathrm{d}t}=v_\mathrm{W} =\frac{\mathrm{d}Q_\mathrm{coll}}{\mathrm{d}t}\frac{1}{\Delta T_\mathrm{W}} \frac{1}{\rho_\mathrm{W}c_\mathrm{W}}\frac{1}{\pi N r^2_\mathrm{W}}\,.
 \label{eq:cool3}
 \end{equation}
From equation~\ref{eq:cool1} follows that a maximum water flow rate of  about 4\,l/s is required. Keeping the values of $v_\mathrm{W}$ at about 1 to 2\,m/s and using $r_\mathrm{W}= 4\,$mm,  22 cooling channels should be placed into the copper jacket encasing the collimators with 9\,cm diameter, and 32  cooling channels  into the copper jacket encasing the collimators with 14\,cm diameter. So the temperature of the cooling water is increased by about 5--10\,K depending on $E_\mathrm{cm}$ and the required luminosity.   The precise numbers for each collimator part can be calculated based on the energy deposition given in table~\ref{tab:CollPar} and using equation~(\ref{eq:cool5}).

The Reynolds number,
\begin{equation}
Re=
\frac{ 2r\langle v_\mathrm{W}\rangle } {\nu \mathrm{(T=30^\circ C)}_\mathrm{kin.vis.}}
\label{eq:reynolds}
\end{equation}
is 11,430 for $v_\mathrm{W}=1\,$m/s indicating turbulent flow; $\nu_\mathrm{kin.vis.}$ is the kinematic viscosity.
The parameter values of water used for these considerations are summarized in table~\ref{tab:water}.

\begin{table}[h]
 \renewcommand{\arraystretch}{1.14}
\centering
\begin{tabular}{|l|c|c|c|}
\hline
 parameter   & $c_\mathrm{W}$ & $\rho_\mathrm{W}$ & $\nu_\mathrm{kin.vis.}$ \\
\hline 
value        & 4182 J/(kg K) & 0.9982 g/cm$^3$ & 7.98437$\times 10^{-7} \,$m$^2$/s  \\
\hline
\end{tabular}
\caption{\label{tab:water}Thermal parameters of water.}
\end{table}

The  heat flux from copper through  the  surface  of $N$ water channels, $N A_\mathrm{W}=2N\pi r_\mathrm{W} L$, corresponds to  
 \begin{equation}
  \frac{\mathrm{d} Q_\mathrm{coll}}{\mathrm{d}t} =\alpha_\mathrm{W} N A_\mathrm{W} \Delta T_\mathrm{W}\,,
 \label{eq:cool4}
 \end{equation}
where $\alpha_\mathrm{W}$ is the heat transmission coefficient to the water and $L$ the length of the cooling channels. The design presented here requires values of $\alpha_\mathrm{W}$ between $0.1\,$W/cm$^2$/K and  $0.6\,$W/cm$^2$/K for a water temperature drop of $10\,$K.    
 
More complex is the heat transfer from the collimator material, \ie\ graphite, titanium, iron, to the copper jacket. In particular, the graphite -- copper connection is important. 

The heat transfer depends strongly on the the surface roughness and the contact pressure. In reference~\cite{ref:DumpXFEL} the heat transfer coefficient from graphite to copper is estimated depending on the contact pressure and the gas filling the gap at the material junction. Based on this considerations a transfer coefficient $\alpha_\mathrm{C\rightarrow Cu}$ of  $0.4\,$W/cm$^2$/K can be achieved.
Taking into account the heat transfer from graphite to copper, 
the difference between the temperatures in copper  at the cooling channels and the inner surface of the collimator is
\begin{equation}
\displaystyle{T_\mathrm{C} - T_\mathrm{Cu}= R_\mathrm{heat}\frac{\mathrm{d}Q}{\mathrm{d}t}}\,,
\label{eq:cool5}
\end{equation}
with the thermal resistance 
\begin{equation}
 R_\mathrm{heat} = \frac{1}{2\pi L}\left[
 \frac{1}{\lambda_\mathrm{C}} \ln\frac{r^\mathrm{C}_\mathrm{a}}{r^\mathrm{C}_\mathrm{i}}
+\frac{1}{r_\mathrm{a}^\mathrm{C}\alpha_\mathrm{C\rightarrow Cu}}
+\frac{1}{\lambda_\mathrm{Cu}} \ln\frac{r^\mathrm{Cu}_\mathrm{a}}{r^\mathrm{C}_\mathrm{a}}\right]\,,
\label{eq:cool6}
\end{equation}
where $r^\mathrm{C}_\mathrm{i}$ is the aperture radius of the graphite part, $L$ the length,  
the contact C--Cu is located at $r^\mathrm{C}_\mathrm{a}$, the outer radius of the graphite and the inner radius of the copper jacket, and $r^\mathrm{Cu}_\mathrm{a}$ is the effective outer radius of the copper at the cooling channels.  
With  realistic values,  $\alpha_\mathrm{C\rightarrow Cu} = 0.4 (0.1)\,$W/cm$^2$/K, 
the average temperature at the inner collimator surface increases by about 15\% (60\%) in comparison to the ideal case neglecting the thermal resistance at the Cu-C junction. This can be accepted since graphite stands substantially higher temperatures than the average temperatures of about 300--400$^\circ$C. 
At the Cu-Ti and Cu-Fe contacts, 
thermal transfer coefficients of 0.4\,W/cm$^2$/K increase the average temperature at the inner surface by about 10\% for the iron and about 2\% for the titanium part
in comparison to the ideal case. Anyhow, the power deposition in the titanium and iron sections is relatively low. 
So the cooling of the whole collimator is not a problem.

\begin{figure}[hp]
  \centering
 \includegraphics*[width=150mm]{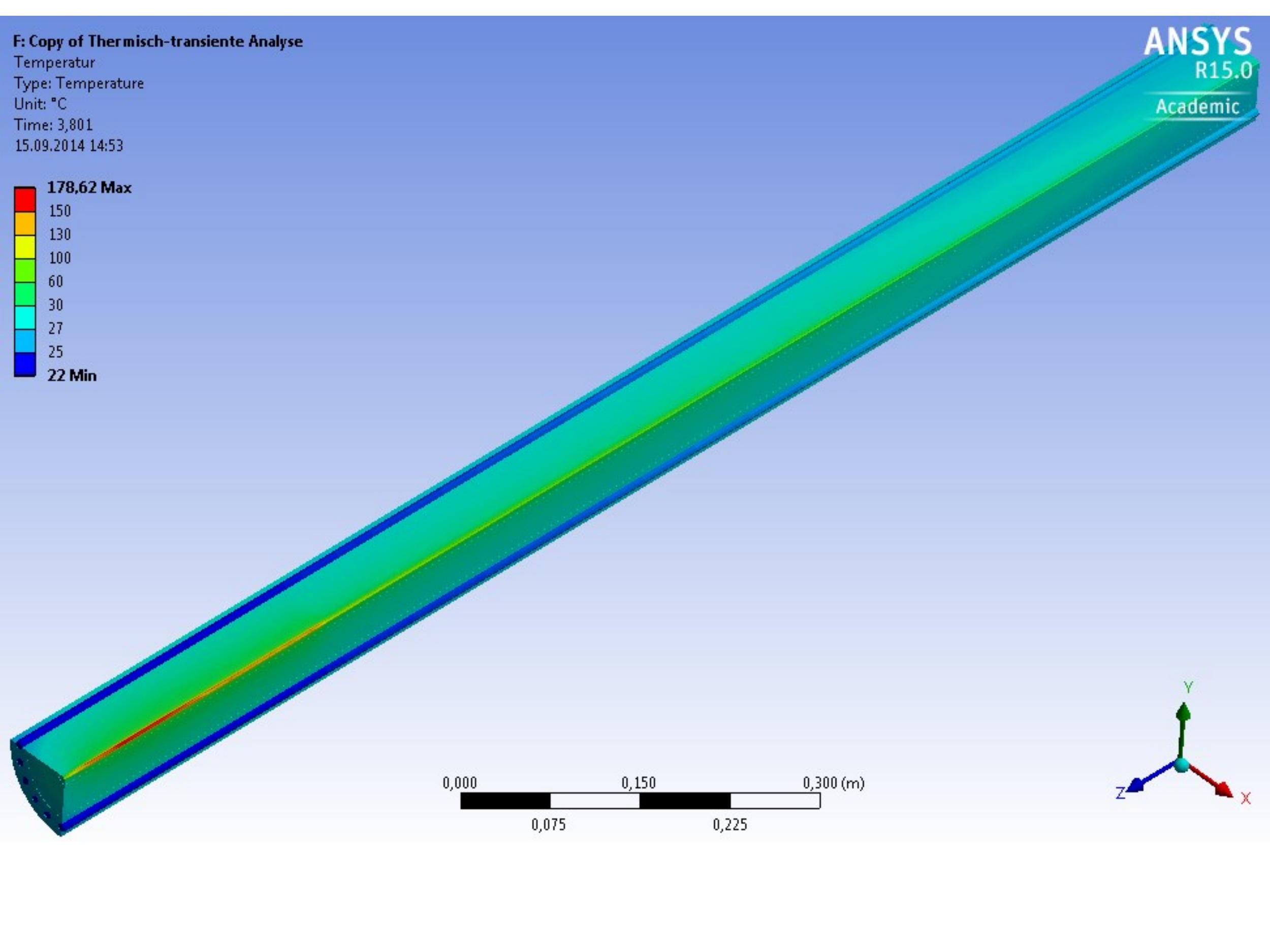}
 \caption{Temperature distribution in the pyrolytic graphite and in the cooling water  for the second  collimator at $E_\mathrm{cm}=350\,$GeV. The figure shows a snapshot at a peak temperature load in the collimator. 
}
\label{fig:CollCoolT}
\end{figure}

It must be remarked that the considerations of the cooling parameters are based on the stationary case and correspond to averaged numbers. Further, 
the temperature distribution is not homogeneous over the collimator, and the temperature of the cooling water depends on its path along the collimator. To get a real picture of the temperature distribution, ANSYS simulations have been performed taking into account the pulsed heating and a water flow of  1\,l/s. 
The resulting temperature distribution for the pyrolytic graphite and the cooling water of the second collimator stage  is shown in figure~\ref{fig:CollCoolT} for $E_\mathrm{cm}=350\,$GeV; the temperature difference of the cooling water is 4.3\,K. 

\section{Load  and potential material degradation 
}\label{sec:problems}

\subsection{Maximum heat load}\label{sec:max-load}
The bunch structure of the ILC beam yields a cyclic load of the collimator material. The maximum values are obtained at the innermost  part of the collimator.
In the chosen collimator design, the instantaneous heating  of pyrolytic graphite by one bunch train reaches  maximum values  of about 124\,K; the maximum heating by one bunch only is below 0.12\,K. 
Since the  heat dissipates, 
the  values for the peak energy density and the corresponding maximum temperature rise due to bunch train are  about 20\% lower than the value expected by multiplying the number of bunches with  the maximum temperature rise by one bunch.
The effect of heat dissipation during one bunch train  is even  less important in the Ti and Fe parts of the collimator since the 
shower particles are spread to a wider region. 

The peak energy deposition per bunch train, the corresponding maximum temperature as well as further important parameters of the power deposition in the collimator are summarized in table~\ref{tab:CollPar} for the different centre-of-mass energies and the collimator parts.  
 
\begin{figure}[hp]
  \centering
   \includegraphics*[width=150mm]{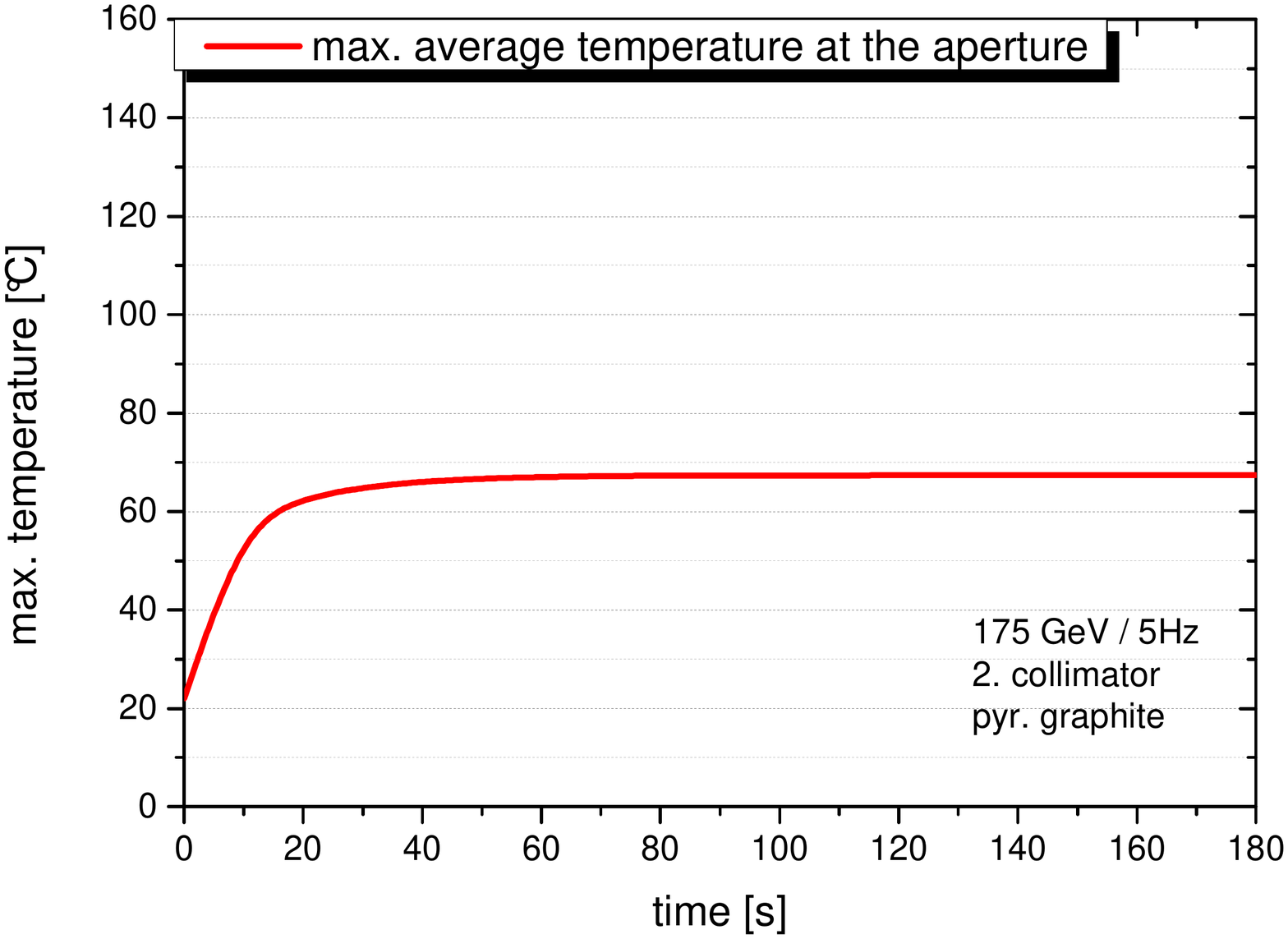}\\
   \includegraphics*[width=150mm]{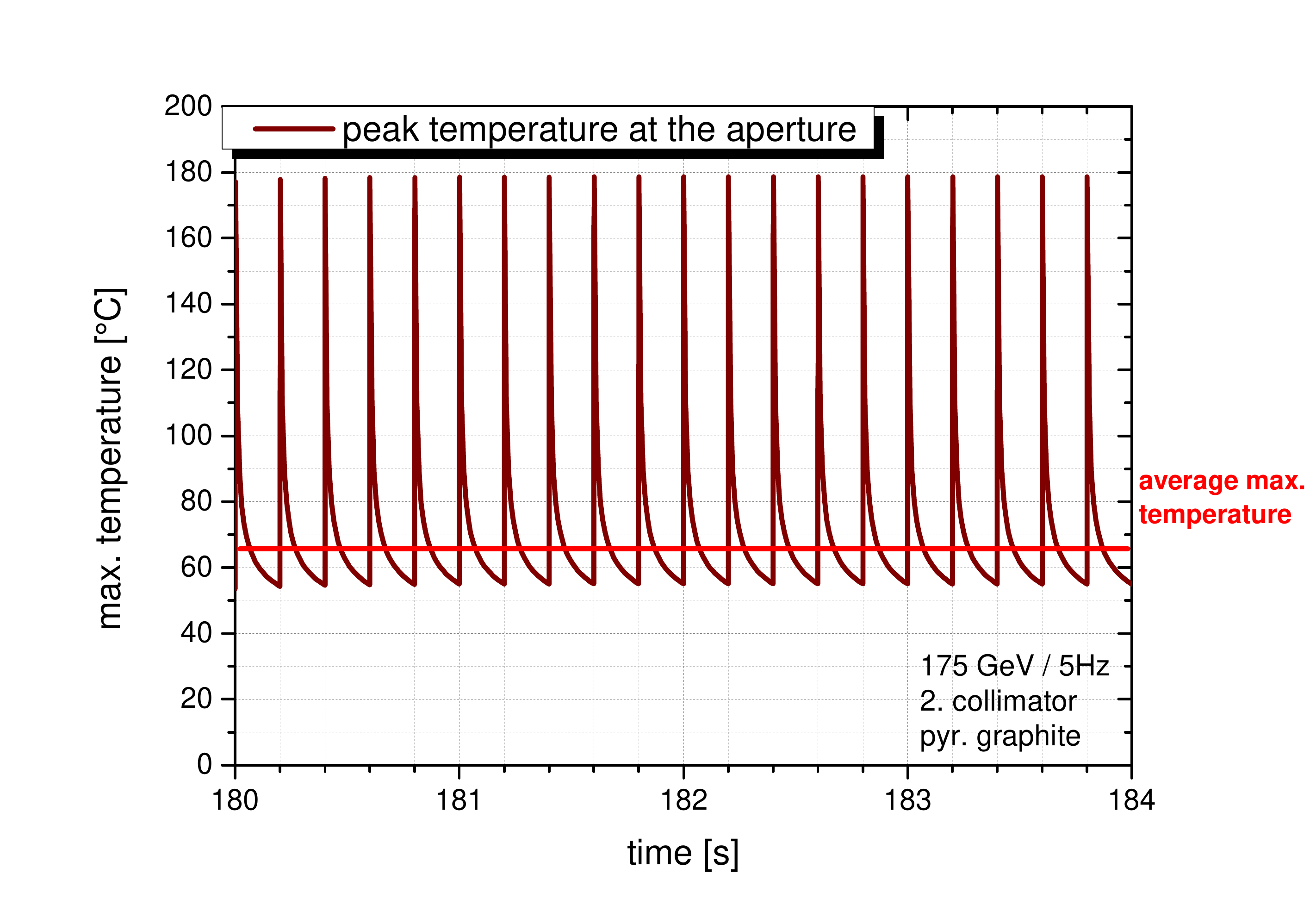}
 \caption{Time-dependent evolution of the  average temperature in the pyrolytic graphite  of the second collimator at the aperture which is  the area with the highest heat load. The upper plot shows the average temperature as function of time for the inner part of the collimator if $E_\mathrm{cm}=350\,$GeV ($E_\mathrm{e^-}=175\,$GeV). The lower plot shows the corresponding temperature evolution  in detail. 
}
\label{fig:Ttime}
\end{figure}

To illustrate the time-dependent temperature evolution  over  bunch trains, figure~\ref{fig:Ttime}  shows the maximum temperature in  the pyrolytic graphite part of the second collimator  for $E_\mathrm{cm}=350\,$GeV.  After several bunch trains the 
average temperature is reached  
which is determined by the collimator dimension, the deposited energy and the  cooling system (see  also equation~(\ref{eq:dQdtcyl})).

\subsection{Stress}\label{sec:pressure}

The rapid energy deposition during one bunch train causes stress inside the collimator since the material is not able to expand as fast as it is heated.
Assuming short, intense beam pulses,  the irradiated zone is  instantaneously heated  under constant volume, leading to a change of pressure; \ie\ the material is in a hydrostatic state of stress.  The corresponding thermo-elastic peak stress value is given by 
\cite{ref:sievers,ref:timoshenko} 
 \begin{equation}
 \Delta \sigma_\mathrm{max} = \frac{Y  \alpha \Delta T}{1-2\nu}\,,
 \label{eq:press}
 \end{equation}  
 where  $\Delta T$ is the the temperature rise per bunch train; the Young's modulus of elasticity, $Y$, the coefficient of thermal expansion, $\alpha$, and Poisson's ratio, $\nu$, can be taken from table~\ref{tab:matpar}.
From the heated zone stress waves emanate. 
However, for the collimator design presented here
the instantaneous temperature rise is too small 
to create dangerous stress waves in the material.  
For example,  the highest instantaneous temperature rise  in the pyrolytic graphite amounts to  roughly 0.1\,K per bunch yielding  about 125\,K per bunch train (1\,ms).
Since pyrolytic graphite is a highly anisotropic material with different  Poisson's ratio and  thermal expansion  in (x,y) and (z) direction,   equation~(\ref{eq:press}) cannot applied to estimate the peak stress in the graphite parts of the collimator. For the parameter values considered here, the maximum stress  can be approximated with
 \begin{equation}
 \Delta \sigma_\mathrm{max} = \frac{Y  \alpha_\mathrm{(z)} \Delta T}{1-2\nu_{\mathrm{(x,y)}}}\,.
 \label{eq:press-appr}
 \end{equation} 
The instantaneous maximum temperature rise of 125\,K in the pyrolytic graphite results in peak stress of about 14\,MPa.

It should be remarked that the material parameters depend on the temperature. The stress development depends strongly on the  temperature, and  equations~(\ref{eq:press}) or (\ref{eq:press-appr}) allow only a rough estimate.

\begin{figure}[h]
  \centering
  \includegraphics[width=150mm]{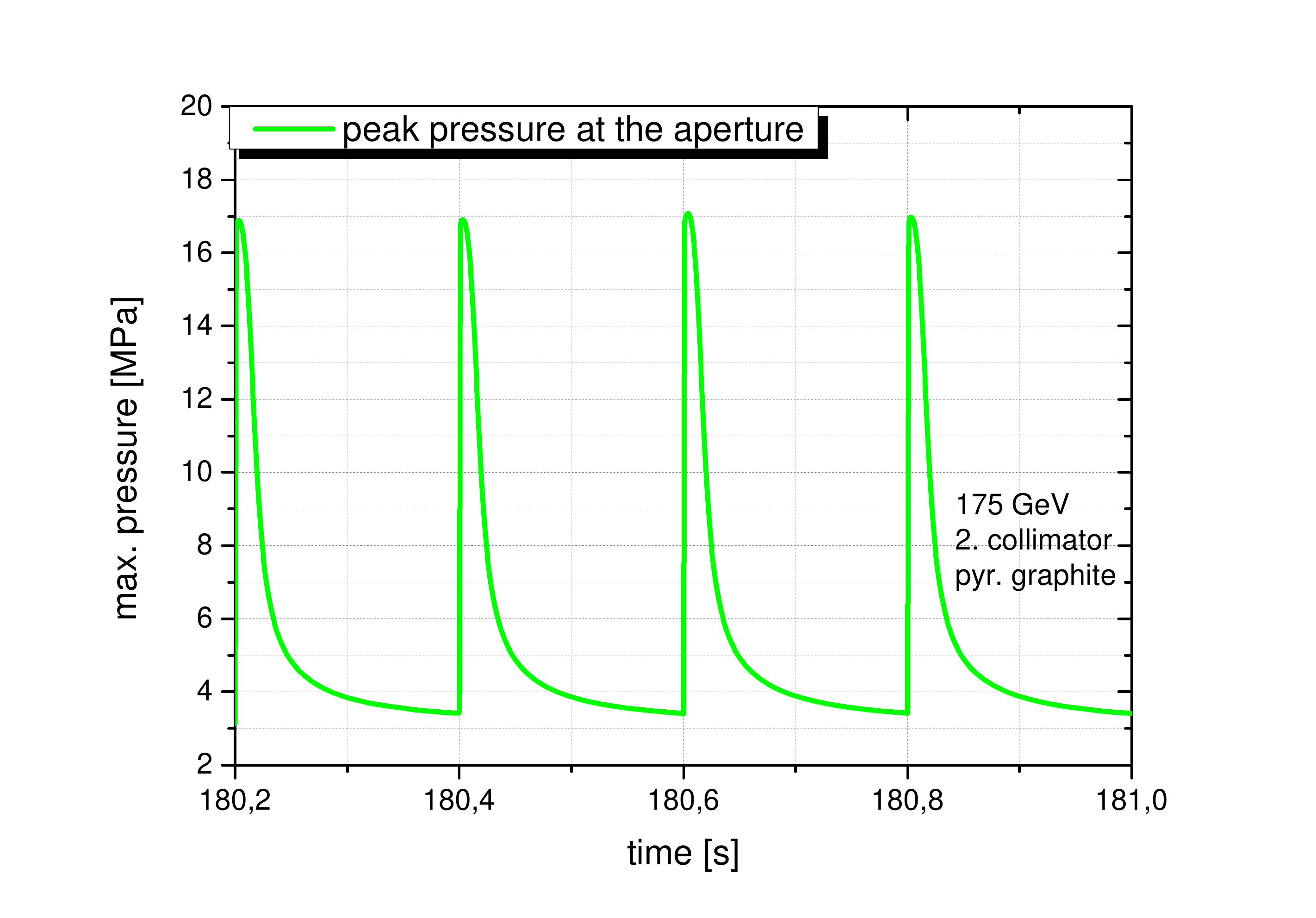}
 \caption{Evolution of the maximum pressure at the aperture in the pyrolytic graphite  of the second collimator for  $E_\mathrm{cm}=350\,$GeV ($E_\mathrm{e^-}=175\,$GeV). 
 }
\label{fig:stress-btrain}
\end{figure}

The peak stress values appear near the collimator aperture, in the hottest region. Since the material cannot expand during  the short time of one bunch-train,   pressure at the inner collimator surface is produced. The time evolution of the maximum pressure at the aperture  in the pyrolytic graphite is visualized in figure~\ref{fig:stress-btrain} for the second collimator for  $E_\mathrm{cm}=350\,$GeV.  Similar distributions have been calculated for all collimator parts and centre-of-mass energies. 
The gradient between the average temperatures of inner and outer collimator region causes a   permanent static stress of few MPa.

The stress evolution is important to evaluate the load  during  long-term operation.
Assuming a  running time of 5,000 hours, the ILC collimator system undergoes   $9\times 10^{7}$ load cycles. However, cyclic load 
could damage the material already at values substantially below the load limit:
A rule of thumb gives about 40\% of the tensile yield strength as fatigue limit. These limits are derived in tests with mechanical load.
At the collimator, the impact of high-energy photons and secondary particles may change the properties of the material and reduce the fatigue limit further. In addition, all material parameters depend on the temperature. If for instance the thermal transfer coefficient is reduced due to long-term irradiation, the average temperature increases and the corresponding fatigue limit could decrease. 
Thus, it is necessary to have a good safety margin.

A comparison of the stress created per bunch train
with the 
parameters given  in table~\ref{tab:matpar} indicates that the cyclic  amplitude does not reach the fatigue limit allowed for the collimator material. However, in case of $0.7\,$mm iris radius to achieve almost 60\% polarization at $E_\mathrm{cm}=500\,$GeV, the peak stress  is increased by a factor 3 (up to 4) and comes close to or  exceeds the fatigue limit.

\subsection{Misalignment}\label{sec:misalign}

For an ideally positioned photon beam the energy deposition in the collimator is below the fatigue pressure limit. However, already a transverse displacement of 100\,$\mu$m increases substantially the energy deposition in the collimator as shown in table~\ref{tab:CollPar} for the high luminosity option at $E_\mathrm{cm}=500\,$GeV. 
For an iris with $r\ge 1\,$mm, the maximum values exceed neither the fatigue limit nor the yield strength. However, the safety margin is reduced.

\subsection{Damage and deformation}\label{sec:damage} 

Since the photon collimator is a dump for a large part of the photon beam, radiation damages of the collimator material must be taken into account. A rough measure of this damage is  the displacement per atom (dpa). These dpa values were determined by FLUKA simulations for each collimator part. 
Table~\ref{tab:dpa} summarizes the  maximum dpa values induced 
for different centre-of-mass energies. Only at the region near the inner collimator surface these high values are obtained; dpa values  decrease in radial direction corresponding to the energy deposition shown in  figure~\ref{fig:ILC_collimatorR12}. 

\begin{table}[h] 
 \renewcommand{\arraystretch}{1.14}
\begin{center}
\begin{tabular}{|l|cc|c|c|} \hline  
      $E_\mathrm{cm}$     & \multicolumn{2}{c|}{125\,GeV} 
                        & 350\,GeV
                        & 500\,GeV     \\\hline
      $E_\mathrm{e-}$     & 125 GeV & 150 GeV
                        & 175 GeV
                        & 250 GeV     \\
                        & (1. collimator) & (1. collimator) 
                        &  (2. collimator)
                        &  (3. collimator)    \\\hline
    pyr. C      &  $1.0$  & $1.6$ & $2.7$ & $3.0$ \\
    Ti8Mn       &  $0.7$  & $1.1$ & $2.2$ & $2.5$ \\
    Iron(St-70) &  $0.5$  & $0.8$ & $1.7$ & $1.7$ \\
 \hline
\end{tabular}
\caption[dpa of the different photon collimator parts and energies in 5000h.]{ \label{tab:dpa}Maximum dpa 
values in the collimator material simulated with FLUKA for
different electron beam energies. } 
\end{center}
\end{table}

It is not easy to find a clear statement up to which dpa levels material can be explored and how the material properties change. In general, titanium alloys and iron should stand values up to 1\,dpa and the highly affected zone is thin. Therefore, one year operation time is probably at the limit. It is recommended to test the material degradation in an experiment before fixing the final design. 

Graphite shows depending on  dpa value and temperature a substantial dimensional change as reported in references, \eg~\cite{ref:simos,ref:swelling}. 
The review~\cite{ref:PyrGraph1} 
describes the swelling and the  change of parameters of pyrolytic graphite  when irradiated by electrons. Depending on dpa and temperature, the material could expand 
in longitudinal and tighten 
in basal direction. 
At the photon collimator, this  dimensional change would appear  at the innermost layer after a longer time of irradiation.  
The size of the dimensional change depends on the temperature and dpa value and also on the special material.  

At a first glance the expansion in longitudinal direction can be accepted since the pyrolytic collimators are made of slices which can be positioned in a safe distance. The expansion in basal direction yields a slight increase of the collimator aperture. This could affect the degree of positron polarization but taking into account the jitter and  of a realistic beam, this effect should be negligible.  
Another problem could arise concerning the stored energy release of irradiated  graphite. Following reference~\cite{ref:stored-energy}, that the amounts of stored energy would be small.

Nevertheless, these problems need further 
studies; in particular concerning long-term stability in case of misalignment.

\subsection{Activation}\label{sec:activation}

\begin{figure}[h]
\centering
\begin{tabular}{lr}
\vspace{-0.5cm}

  \includegraphics[width=0.45\textwidth]{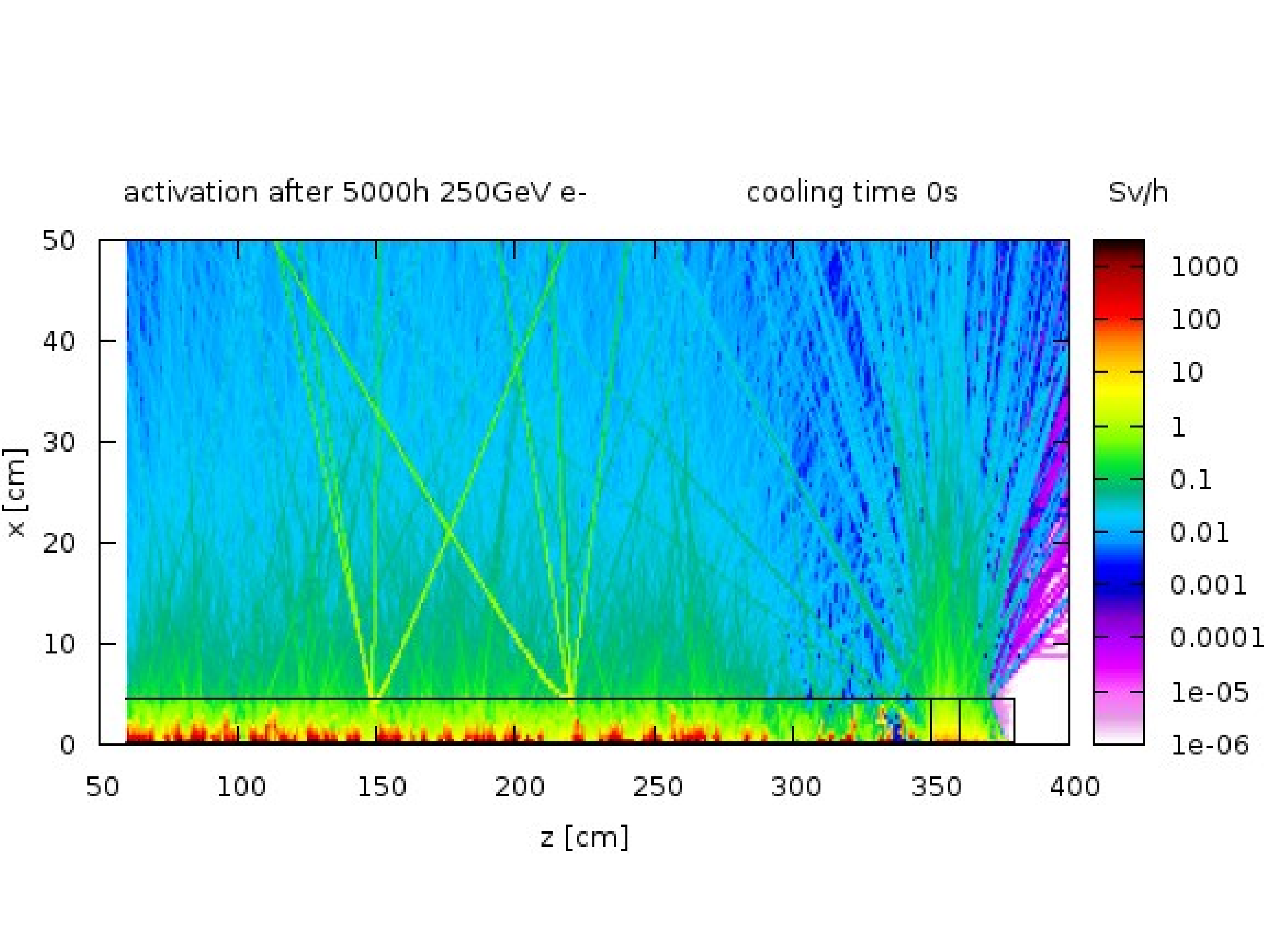}&
  \includegraphics[width=0.45\textwidth]{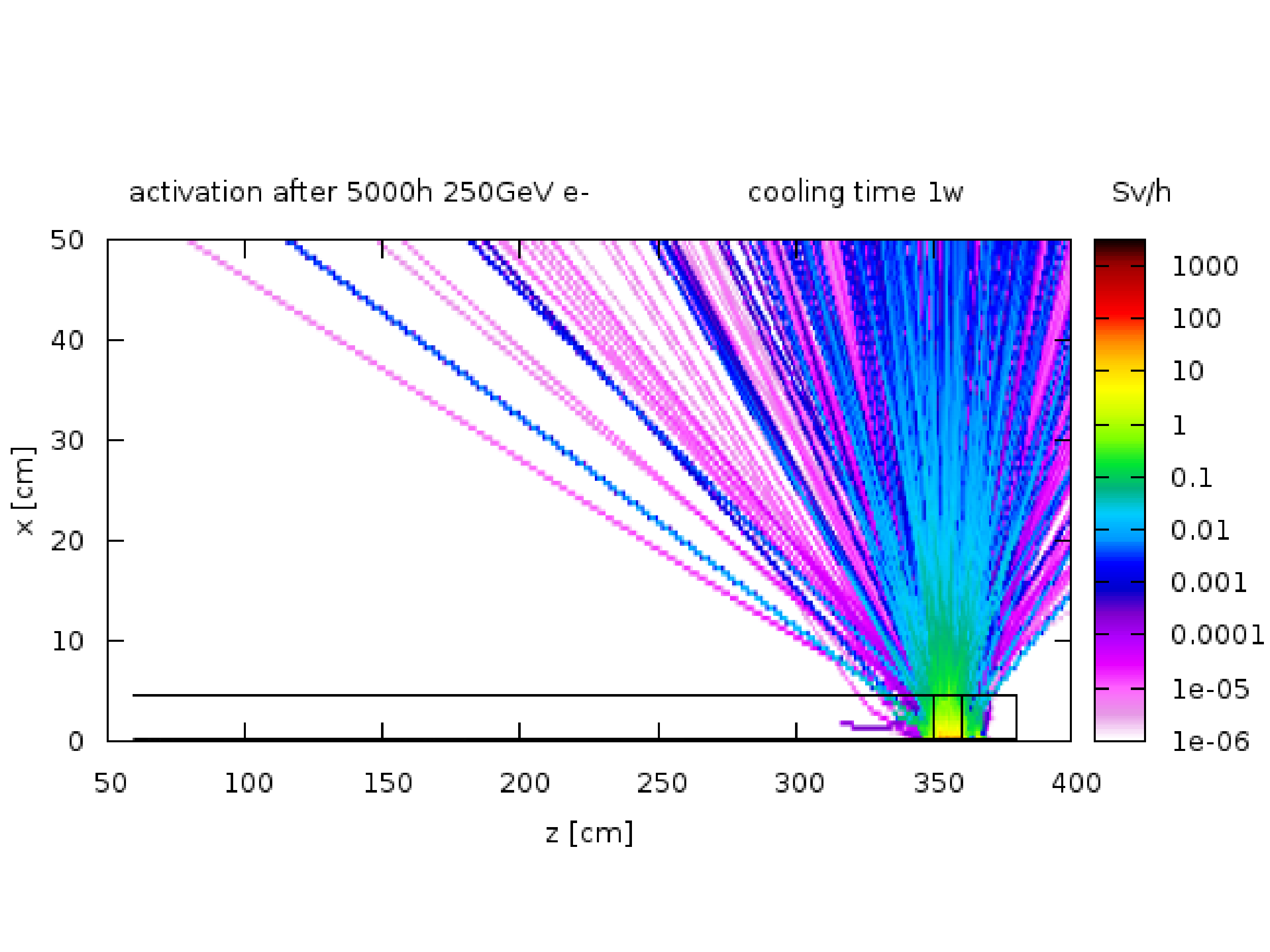}
\end{tabular}
 \caption{Equivalent dose at the third  collimator for $E_\mathrm{cm}=500\,$GeV. 
The left plot shows the equivalent dose 0\,h after  5000\,h irradiation time;  the right plot after one week cooling time. }
\label{fig:coolingtime}
\end{figure}

The collimator material is exposed to a high radiation dose and nuclear reactions are triggered. 
The activation induced by  the photon beam and the secondary particles including neutrons has been calculated using the FLUKA Monte Carlo code for particle tracking and particle interactions with matter \cite{ref:FLUKA}. In  figure~\ref{fig:coolingtime}, the equivalent dose after 5000 hours operation  is shown  up to a radial distance of 0.5\,m  for the third collimator at $E_\mathrm{cm}=500\,$GeV. %
After one week cooling time only the long-living nuclei contribute to the equivalent dose.  The highest dose 
comes from 
the aperture of the Ti8Mn part; at a distance of  0.5\,m the equivalent dose amounts roughly 30\,mSv/h. Following the half-life values given in table~\ref{tab:decay1w},  this value reduces only slowly. 
 Therefore, 
the activated components must  be handled with special care.

\begin{figure}[hp]
\centering
\begin{tabular}{lr}
  \includegraphics[width=0.45\textwidth,height=5cm]{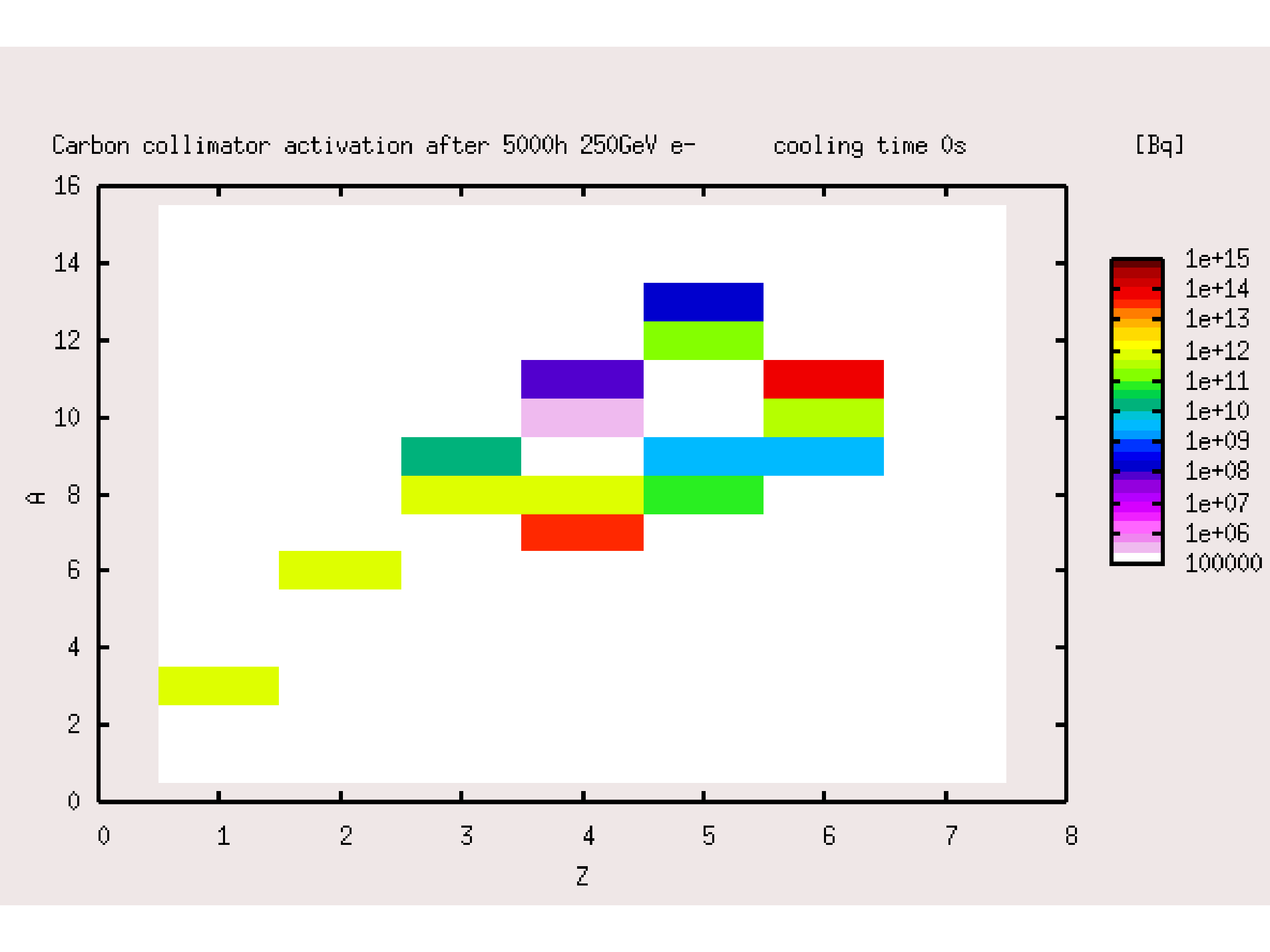}&
  \includegraphics[width=0.45\textwidth,height=5cm]{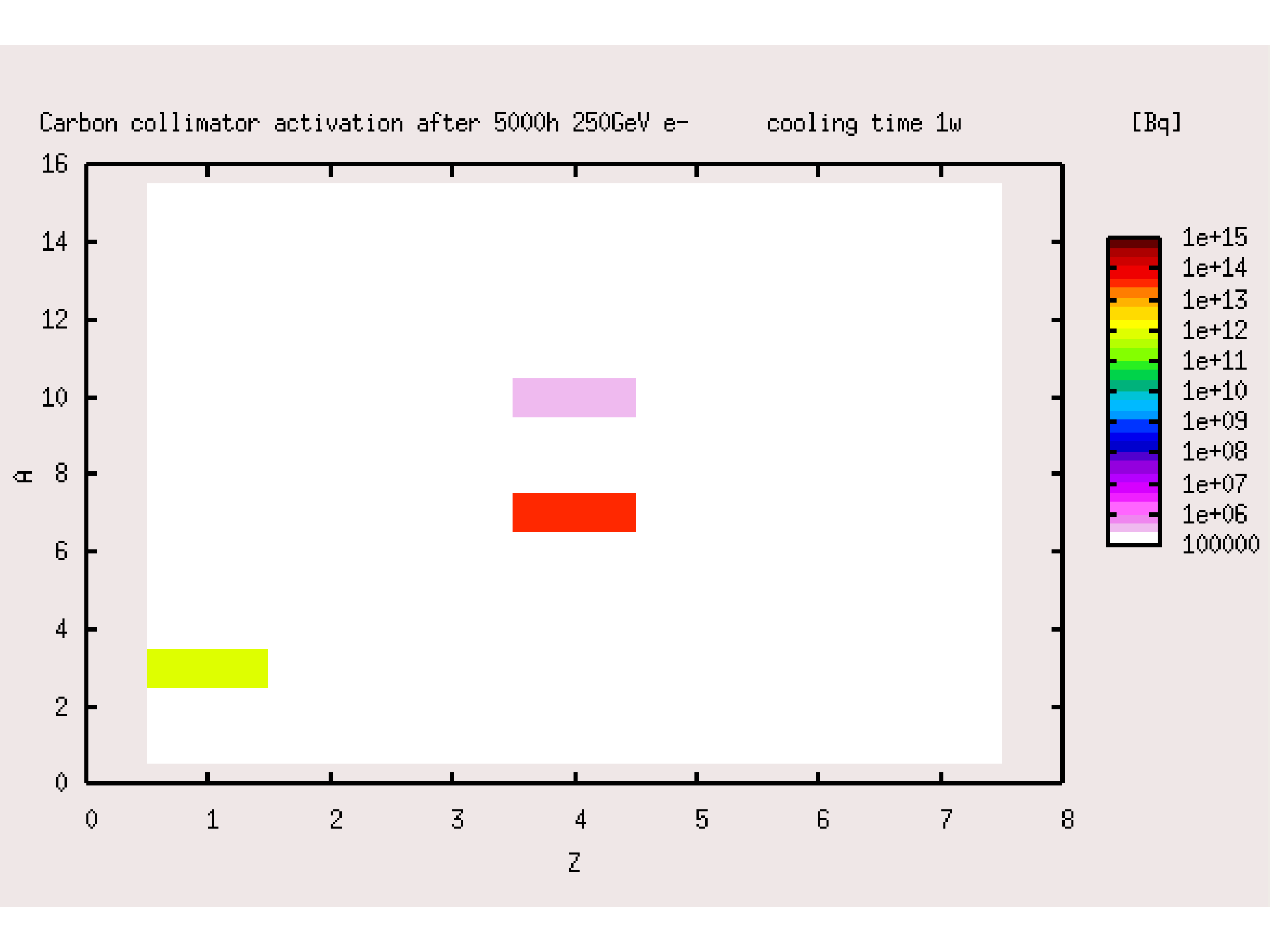}\\
  \includegraphics[width=0.45\textwidth,height=5cm]{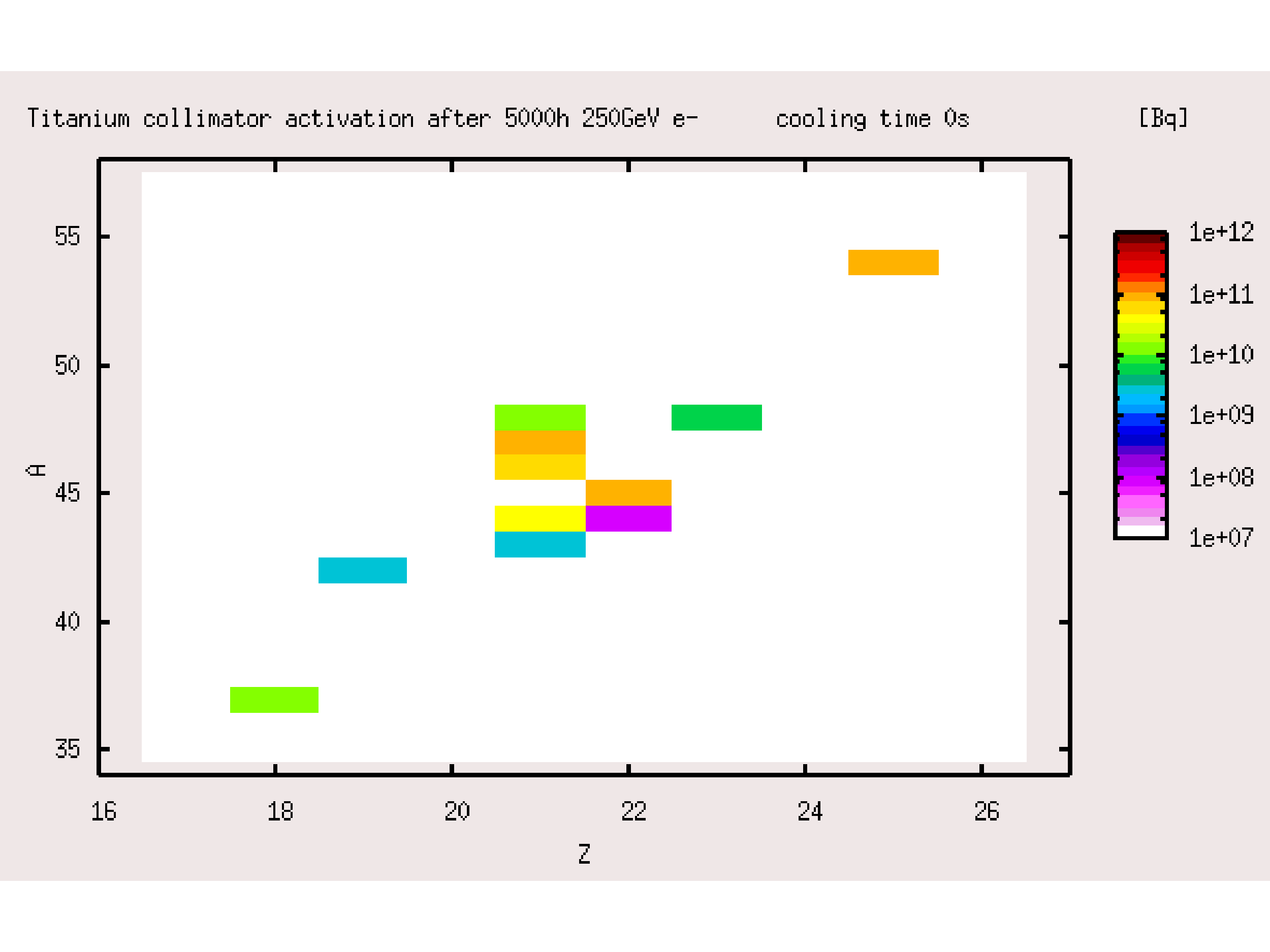}&
  \includegraphics[width=0.45\textwidth,height=5cm]{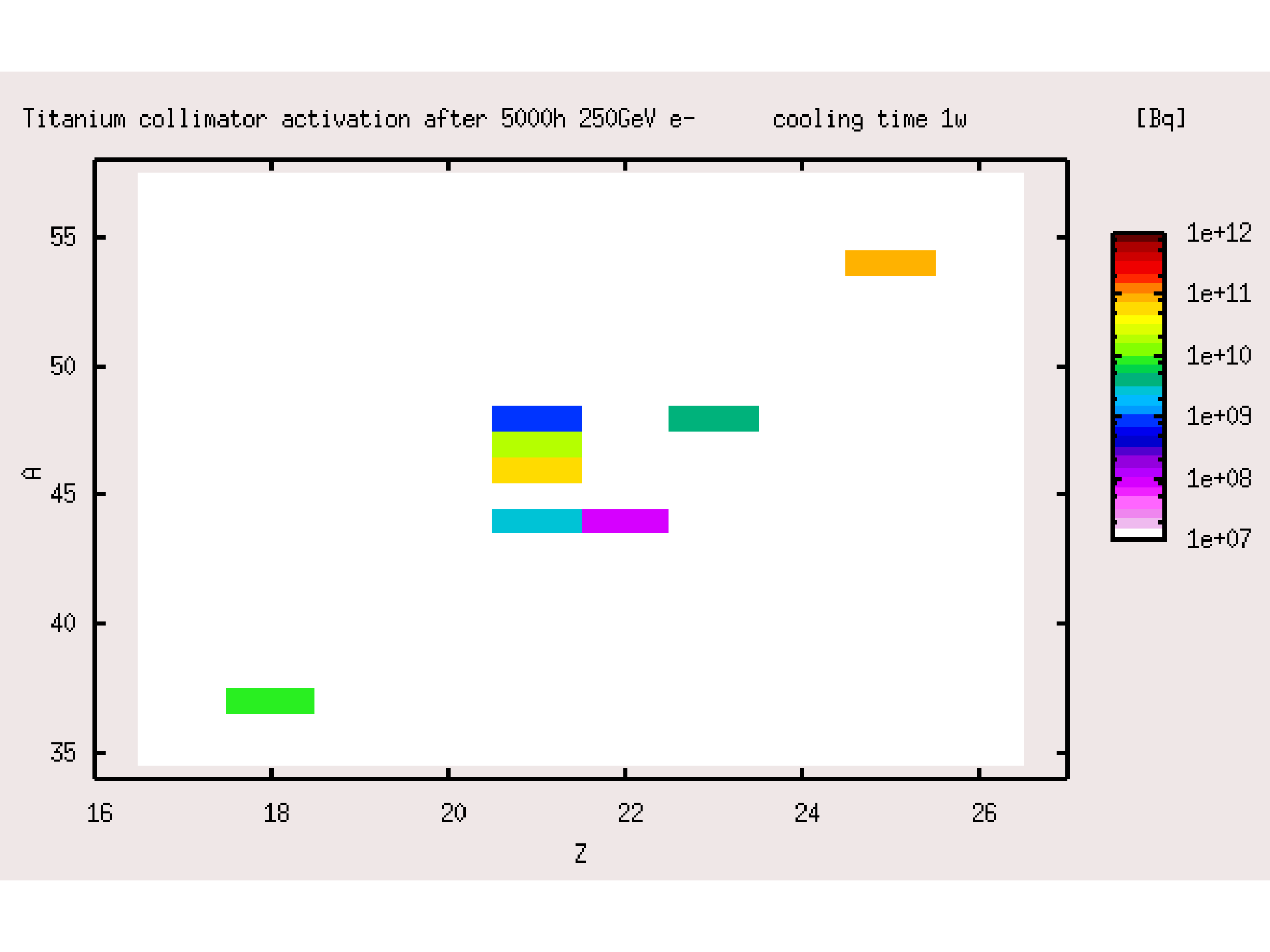}\\
  \includegraphics[width=0.45\textwidth,height=5cm]{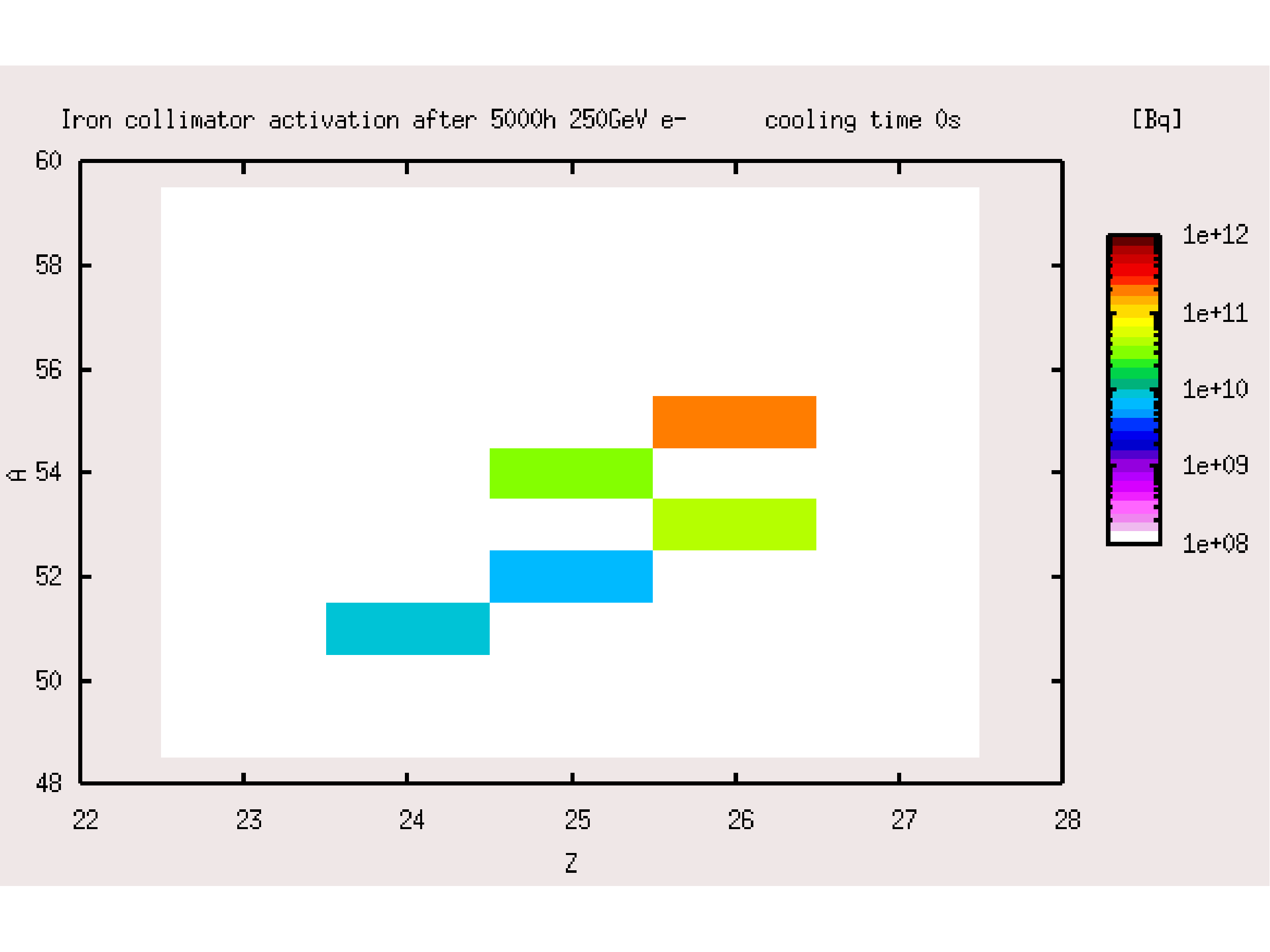}&
  \includegraphics[width=0.45\textwidth,height=5cm]{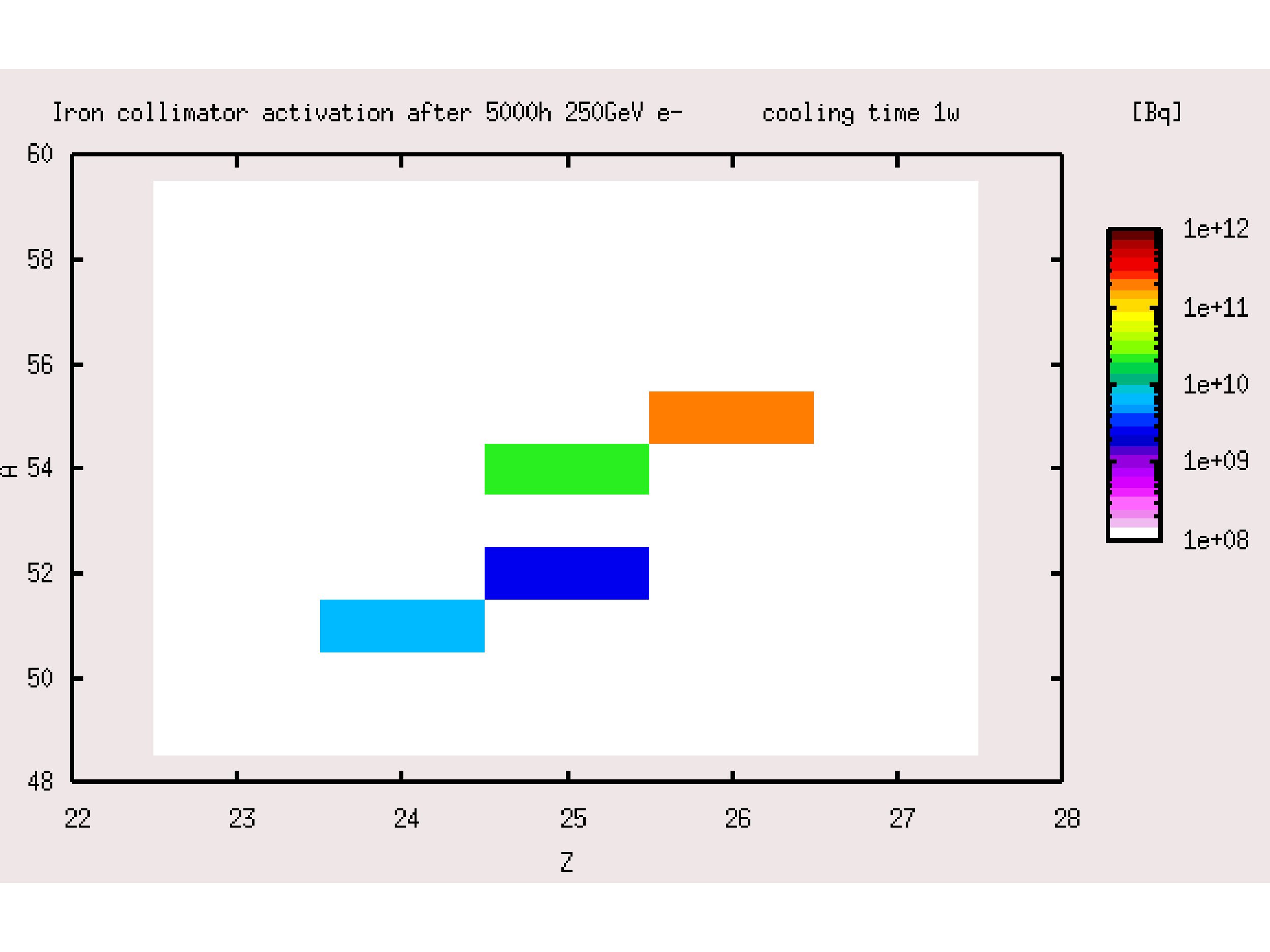}\\
\end{tabular}
 \caption{\label{fig:coolingelements}Activation in the pyrolytic graphite, titanium and iron part of the third collimator for $E_\mathrm{cm}=500\,$GeV.  The left plots show the activated nuclei --given by atomic number $Z$ and atomic weight $A$--  0\,h after irradiation, the right plots after one week cooling time.}
\end{figure}

To illustrate the activation in the collimator, figure~\ref{fig:coolingelements} shows the the isotopes given by the atomic number $Z$ and the atomic weight $A$ and the decay rates after 5000 hours irradiation and after one week cooling time. 
%

 In table~\ref{tab:decay1w} the nuclei with significant activity produced  in the collimator material are listed. 

\begin{table}[hp]
 \renewcommand{\arraystretch}{1.14}
\centering
\begin{tabular}{|l|c|c|c|}
\hline
  \multicolumn{4}{|c|}{pyrolytic graphite}  \\
\hline 
nucleus        & half-life period & activation after 0\,s [Bq] & activation after 1\,w [Bq]  \\
\hline
$^3_1H$       & 12.33 y & $7.0\times 10^{11}$ & $7.0\times 10^{11}$   \\
\hline
$^7_4Be$       & 53.3 d & $2.4\times 10^{13}$ & $2.2\times 10^{13}$  \\
\hline
$^{10}_4Be$       & $1.5 \times 10^6 $y & $4\times10^5$ & $4\times10^5$  \\
\hline
$^{11}_6C$       & 20.39 min & $5.9\times 10^{13}$ & -  \\
\hline\hline
  \multicolumn{4}{|c|}{Ti8Mn}  \\
\hline 
nucleus        & half-life period & activity after 0\,s [Bq] & activity after 1\,w [Bq]  \\
\hline
$^3_1H$       & 12.33 y & $1.6\times 10^{8}$ &  $1.6\times 10^{8}$ \\
\hline
$^{37}_{18}Ar$       & 35.04 d & $1.0\times 10^{10}$ & $9.0\times 10^{9}$  \\
\hline
$^{39}_{18}Ar$       & 269 y & $7.5\times10^6$ & $7.5\times10^6$  \\
\hline
$^{45}_{20}Ar$       & 21.5 s & $1.1\times 10^{11}$ & -  \\
\hline
$^{47}_{20}Ca$       & 4.54 d & $4.1\times10^{7}$ & $1.4\times 10^{7}$  \\
\hline
$^{44}_{21}Sc^{(6+/2+)}$       & 58.6 h / 3.93 h & $1.7\times 10^{11}$ & $5.7\times 10^{9}$  \\
\hline
$^{46}_{21}Sc$       & 83.79 d & $6.6\times10^{10}$ & $6.2\times10^{10}$  \\
\hline
$^{47}_{21}Sc$       & 3.35 d & $9.8\times 10^{10}$ &$2.3\times10^{10}$  \\
\hline
$^{48}_{21}Sc$       & 43.67 h & $1.2\times10^{10}$ & $8.0\times10^8$  \\
\hline
$^{44}_{22}Ti$       & 49 y &  $8.9\times10^7$&  $8.9\times10^7$  \\
\hline
$^{48}_{23}V$       & 15.97 d & $5.0\times10^{9}$ &  $3.8\times10^{9}$  \\
\hline
$^{54}_{25}Mn$       & 312.12 d & $7\times10^{10}$ & $7.2\times10^{10}$  \\
\hline\hline
  \multicolumn{4}{|c|}{Fe (St-70)}  \\
\hline 
nucleus        & half-life period & activity after 0\,s [Bq] & activity after 1\,w [Bq]   \\
\hline
$^{51}_{24}Cr$       & 27.70 d & $7.8\times 10^{9}$ & $6.5\times 10^{9}$\\
\hline
$^{52}_{25}Mn$       & 5.59 d & $6.9\times10^{9}$ & $2.9\times10^{9}$  \\
\hline
$^{53}_{26}Fe$       & 8.51 min & $5.8\times10^{10}$ & -  \\
\hline
$^{54}_{25}Mn$       & 312.12 d & $2.6\times 10^{10}$ & $ 2.6\times 10^{10}$  \\
\hline
$^{55}_{26}Fe$       & 2.73 y & $1.8\times10^{11}$ & $ 1.8\times10^{11}$  \\
\hline
\end{tabular}
\caption{\label{tab:decay1w}Nuclei with  significant activity in the third  collimator for $E_\mathrm{cm}=500\,$GeV after 5000 hours irradiation and after one week  cooling time, respectively. The numbers result from simulations using FLUKA.
}
\end{table}

\section{Alternative collimator design}\label{sec:alternative} 
The photon collimator design presented here has disadvantages, in particular  the use of pyrolytic graphite and the   long extension. So it is worth to think about alternative design possibilities. 

An obvious idea would be to replace at least a large part of the pyrolytic graphite by another material, for instance tungsten. The radiation length of tungsten is a factor ~50 smaller than that of graphite, and tungsten resists very high temperatures. However, for the ILC undulator and the given  energy of the electron drive beam, the particle multiplication in the electromagnetic shower causes  large peak energy deposition.  For tungsten the critical energy, \ie\ the threshold that Bremsstrahlung and pair-production dominates the ionization process, is much lower  than for graphite ($E_\mathrm{crit}^\mathrm{W} = 7.97\,$MeV, $E_\mathrm{crit}^\mathrm{C} = 81.74\,$MeV) and it is below the energy of the photons. The shower created in tungsten causes a load  which exceeds the recommended limits.  
This  cannot be avoided by tapered apertures as done for design of the graphite parts: 
Due to the relatively low critical energy, 
the shower maximum in tungsten --and thus the energy deposition maximum-- is about $1-2$ radiation length in the bulk while in graphite the maximum energy deposition is at/near the surface.

A better idea is to create a collimator design with rotating spoilers alternating with absorber material to stop the outer part of the photon beam. Such design is currently under consideration and development~\cite{ref:colli-altern}.

In another idea  the collimator is integrated in the positron target design: A high Z material (large pair production cross section) is used for the  conversion target and embedded in a low Z material (lower pair production cross section). The geometrical dimension, \ie\ the height, of this high-Z target material can be chosen such that it corresponds to the required aperture of a photon collimator. Considering a spinning target, such principal design provides photon beam collimation in y-direction. In order to collimate also in x-direction, jaws can be added in front of the target entrance. The jaws could be moved (up-down) to avoid thermal overload. The idea of such system is illustrated in~\cite{ref:colli-altern}. Further studies are necessary to evaluate whether this scheme could be realized.
But first rough simulations show that already the clever choice of the height of the converter target material increases the positron polarization. Table~\ref{tab:e+pol-alt-colli} gives an overview of the positron polarization which could be achieved.
\begin{table}[hp]
\begin{center} 
\renewcommand{\arraystretch}{1.14}
\begin{tabular}{|lc|c|c|c|}
\hline
$E_\mathrm{cm}$                & [GeV]  & {250} &{350} & {500} \\
$E_\mathrm{e^-}$               & [GeV] &{125} &{175} & {250} \\\hline
target height                & [mm]  &  4   &  2.8    &  2\\ 
$P_\mathrm{e^+}$(target only)  & [\%]  & 37   & 41    & 32 \\
$P_\mathrm{e^+}$(target$+$jaws)& [\%]  &  --    & 60    & 47 \\
\hline
\end{tabular}
\caption{\label{tab:e+pol-alt-colli}Expected positron polarization, $P_\mathrm{e^+}$, for  $K=0.92$  and $\lambda_\mathrm{und} = 11.5\,$mm for different centre-of-mass energies; the positron yield is 1.5\,e$^+$/e$^-$. The height of the photon conversion target determines the degree of e$^+$ polarization which can be further increased by jaws. }
\end{center}
\end{table}

\section{Summary}\label{sec:summary}

A high degree of positron polarization is desired for physics studies and can be achieved by collimating the undulator photon beam.
Due to the close correlation between energy of the electron beam which passes the helical undulator,  photon beam intensity, collimator iris  and  degree of polarization,
the photon collimator system must be flexible. Further, it 
has to withstand huge heat loads without breakdown during a long operation time. 

The multistage  collimator design presented in this paper represents a  solution to collimate  the photon beam  at the ILC positron source.  For centre-of-mass energies up to 500\,GeV, the material loads stay within acceptable limits taking into account  an additionally safety margin against failure due to fatigue stress.
Depending on the centre-of-mass energy, one, two or all three stages are used to collimate the photon beam. The system is water-cooled, the principal parameters of the cooling system are given.  
The presented solution can be adopted to electron  beam energies up to 500\,GeV. 
However, further simulation studies are recommended to optimize the design taking into account the special material properties as swelling of pyrolytic graphite or potential 
change of properties of the material due to long-term irradiation.  This will further improve the reliability of the final design.

\end{document}